\documentclass[letterpaper, preprint, paper,10pt]{AAS}	

\usepackage{bm}
\usepackage{amsmath}
\usepackage{subfigure}
\usepackage[colorlinks=true, pdfstartview=FitV, linkcolor=black, citecolor= black, urlcolor= black]{hyperref}
\usepackage{footnpag}			      	
\usepackage[font=bf]{caption}

\PaperNumber{13-810}

\begin{document}

\title{A HIGH EARTH, LUNAR RESONANT ORBIT FOR LOWER COST SPACE SCIENCE MISSIONS}

\author{Joseph W.~Gangestad\thanks{Member of the Tech.~Staff, The Aerospace Corporation, M4-947, P.O. Box 92957, Los Angeles, CA, 90009},  
Gregory A.~Henning\thanks{Member of the Tech.~Staff, The Aerospace Corporation, M1-013, P.O. Box 92957, Los Angeles, CA, 90009},
Randy Persinger\thanks{Senior Project Leader, The Aerospace Corporation, M1-013, P.O. Box 92957, Los Angeles, CA, 90009}, \\ and
George R.~Ricker\thanks{TESS Principal Investigator, MIT Kavli Institute for Astrophysics and Space Research, 37-501, 70 Vassar Street, Cambridge, MA, 02142}
}

\maketitle{}

\begin{abstract}
Many science missions require an unobstructed view of space and a stable thermal environment but lack the technical or programmatic resources to reach orbits that satisfy these needs. This paper presents a high Earth orbit in 2:1 resonance with the Moon that provides these conditions, reached via lunar gravity assist. Analytical guidance and numerical investigation yielded deep insight into this unconventional orbit's behavior, making it possible to select a robust mission design. Solutions are available for a broad range of missions, from smaller Explorer-class missions such as the Transiting Exoplanet Survey Satellite to larger missions that seek lower-$\Delta V$ alternatives to traditional Lagrange-point, drift-away, and geosynchronous orbits.
\end{abstract}

\section{Introduction}
NASA astrophysics robotic science missions often require continuous, unobstructed fields of view of the celestial sphere and orbits that provide a stable thermal and attitude-control environment. To date, most large-scale missions use the second Earth-Sun Lagrange point (L2) approximately 1.5 million km from the Earth or a ``drift away'' heliocentric orbit to distances greater than 10 million km, as used by the Kepler mission~\cite{Borucki:2005}. Other observatories may consider geosynchronous orbit, which requires large propellant mass and propulsion systems. The high cost associated with reaching these orbits spurred a search for alternatives that could offer the same operational environment for comparatively little spacecraft $\Delta V$. While the work presented here is applicable to a range of missions, from small ($\sim$350 kg) to large ($\sim$7,000 kg) observatories, this search was performed in particular as part of the orbital analysis for the Transiting Exoplanet Survey Satellite (TESS) mission. TESS is an all-sky, two-year photometric exoplanet discovery mission focusing on Earths and super-Earths around close, bright stars.

The mission orbit selected for TESS early in the design process was the ``P/2-HEO,'' a high Earth orbit in 2:1 resonance with the Moon (i.e., having an orbital period of 13.7 days), first studied in depth by McGiffin~\cite{McGiffin:2001} and similar to the 3:1 lunar-resonant orbit of the IBEX mission~\cite{Carrico:2011}. The P/2-HEO is eccentric, with perigee above geostationary altitude (GEO) and apogee beyond the Moon's orbital radius. The spacecraft reaches this orbit via a gravity-assist flyby of the Moon. The P/2-HEO was selected for TESS to provide unobstructed observation sectors and continuous light curves for the TESS all-sky survey. Although TESS was the model for this study, the results contained herein apply to space missions of any size that require a similar operating environment and seek a low-$\Delta V$ option.

The orbit-parameter selection process for the P/2-HEO was driven by three constraints imposed by TESS: 1) no eclipse can exceed 6 hours duration during the 4-year mission; 2) maintain mission-orbit perigee above GEO ($>$6.6 $R_E$) for four years; and 3) maintain mission-orbit perigee below 22 $R_E$ for four years to ensure robust communications. These parameters may be adjusted as desired for other missions of interest. Extensive orbit analyses were performed to optimize the TESS science orbit and to better understand the unique aspects of the P/2-HEO. A combination of analytical guidance, numerical analysis in the form of tens of thousands of batched orbit propagations, and patched, targeted orbit analysis using Systems Tool Kit (STK) were incorporated to investigate the orbit-parameter space in the presence of lunar and solar perturbations. The analysis included patched, high-fidelity STK orbit propagations for 27 days of the June 2017 lunar cycle, which is one candidate launch period for TESS. Once a desired set of 27 daily mission orbits was determined, a single Design Reference Mission (DRM) was chosen. The P/2-HEO can be represented and bounded by a single lunar cycle; however, several different lunar cycles were evaluated to ensure that seasonal dependence was included and understood.

Starting with an analytical treatment of the orbital problem, this paper shows that the trade space for the P/2-HEO can be reduced to only two variables. Assuming the selection of a final elliptical phasing orbit with a perigee (600 km altitude) and apogee (400,000 km), the initial mission-orbit perigee and inclination are the only remaining control variables for a given time of lunar flyby. Mission-related constraints on orbit variability are related to initial mission- and transfer-orbit lunar inclination, which in turn is controlled by the transfer-orbit semimajor axis. Subsequent numerical investigations verify this analytical guidance and establish how far some assumptions may be relaxed and still yield a desirable mission orbit. The DRM for TESS is presented, demonstrating an example case of how this analysis of the P/2-HEO can be applied to yield a functional mission orbit. Lastly the $\Delta V$ for the June 2017 lunar cycle is presented for the TESS phasing-orbit design (nominal first orbit apogee of 250,000 km) as well as for a hypothetical direct injection, via the Atlas V, of a $\sim$7,000 kg observatory into an initial phasing orbit with an apogee of 100,000 km.

\section{The P/2-HEO}
\begin{figure}[h]
\centering
\subfigure[P/2-HEO in Earth-Moon rotating frame.]{\includegraphics[width=2.5in]{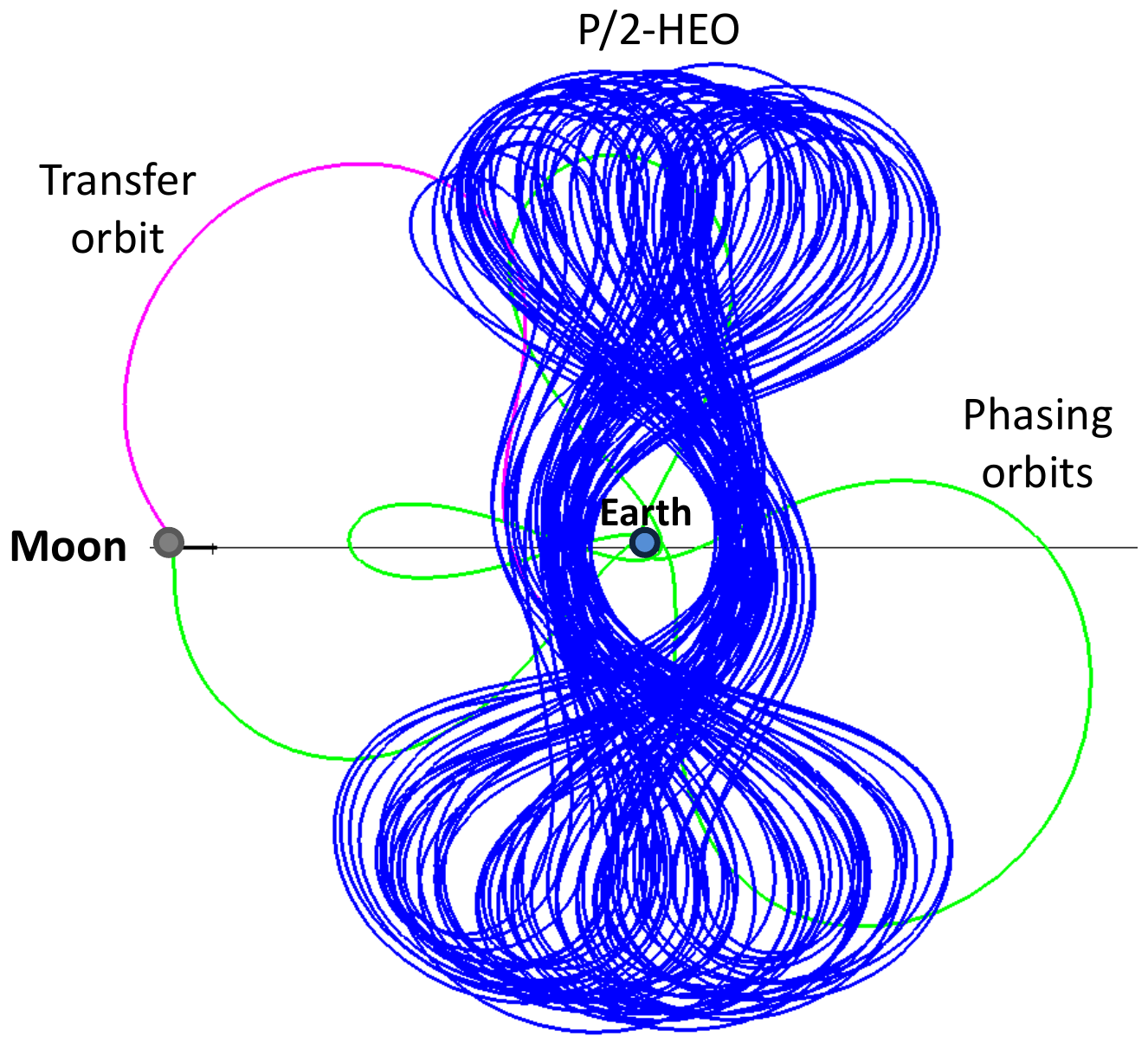}\label{fig:TESSrotTraj}}\qquad
\subfigure[DRM orbital elements, 25 years.]{\includegraphics[width=3in]{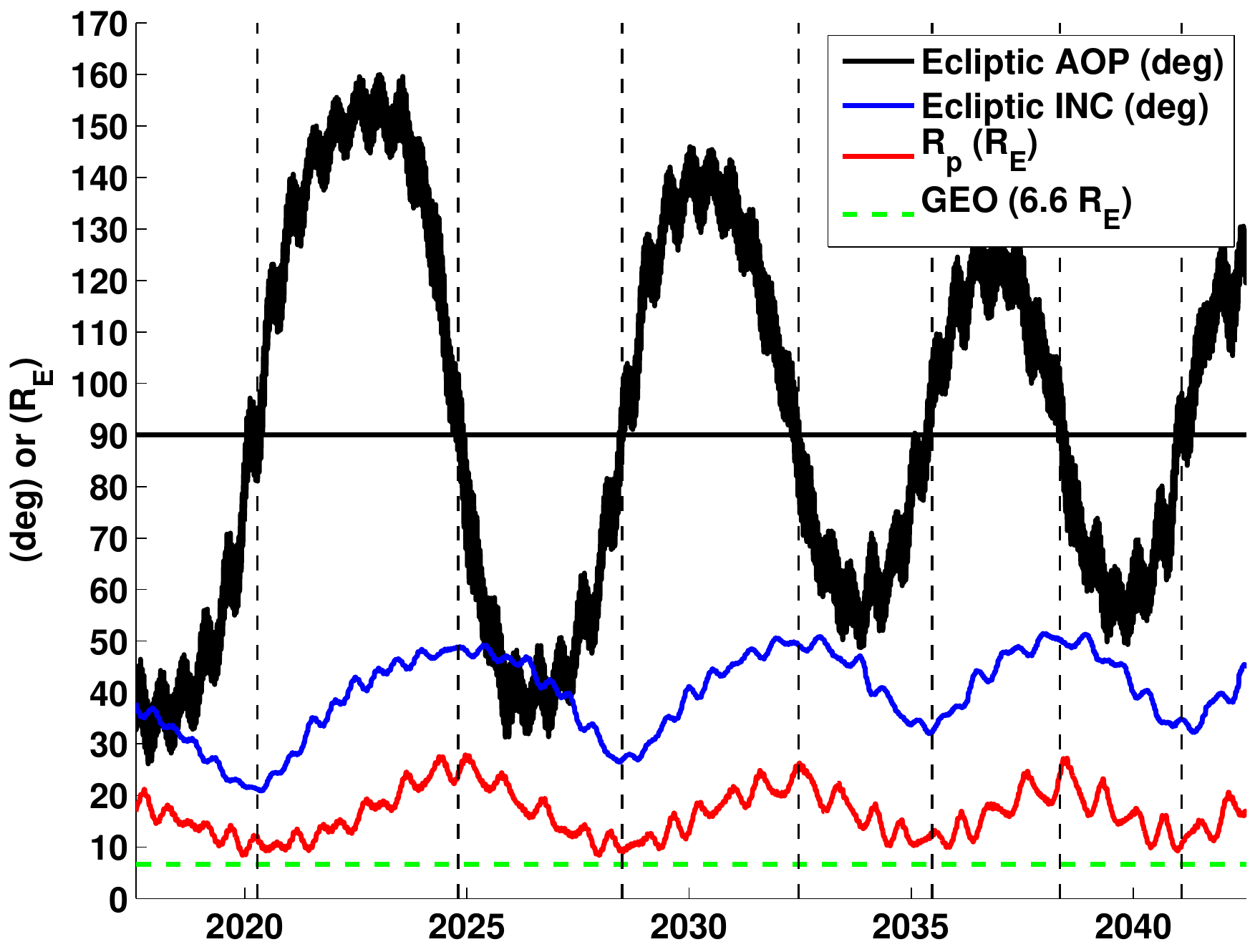}\label{fig:TESSorbels}}
\caption{The P/2-HEO mission, where apogee is offset $\sim$90 deg with respect to the Moon and apogees alternatively lead and trail the Moon. The trajectory plot at left shows the phasing orbits (green), transfer orbit (magenta), and 25-year propagation of the mission orbit (blue). The plot at right shows the orbital elements of the mission orbit as a function of time. Argument of perigee oscillates symmetrically about 90 deg, and all the other orbital elements reach their maxima or minima when the argument of perigee passes through 90 deg, as marked by the dashed lines. }
\label{fig:TESSrot}
\end{figure}

The P/2-HEO analysis draws upon on the initial work of McGiffin~\cite{McGiffin:2001}. A key result of this work is that ``lunar secular perturbations average, roughly, to zero, resulting in significant long term stability'' when the spacecraft apogee is offset $\sim$90 deg with respect to the Moon and when spacecraft apogees alternately lead and trail the Moon, as shown in the Earth-Moon rotating coordinate system in Fig.~\ref{fig:TESSrotTraj}. Figure~\ref{fig:TESSrot} shows the 3.5 phasing orbits, the transfer orbit, and the 25-year propagation of the P/2-HEO mission orbit, including plots of the orbital elements. Although the orbit is high and subject to lunisolar perturbations, a proper selection of initial conditions---in conjunction with resonance with the Moon---drives the orbital elements of the P/2-HEO to oscillate rather than grow or shrink without bound. Much of the subsequent analysis focuses on finding those initial conditions that ensure such non-secular behavior and ensure that the oscillations are bounded by the mission's eclipse and perigee-range constraints.

\begin{figure}[h]
\centering
\includegraphics[width=3in]{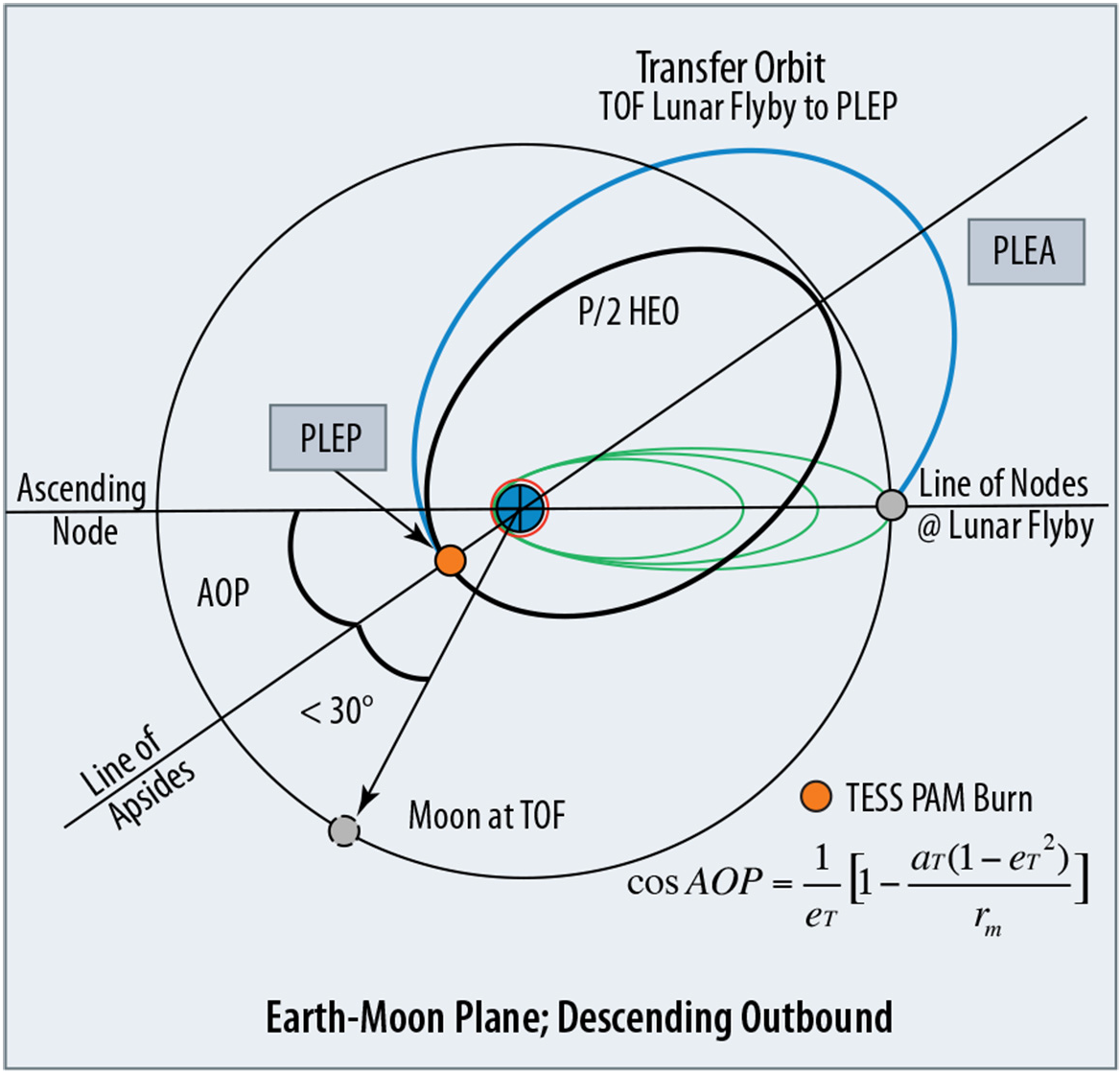}
\caption{The complete P/2-HEO mission, from launch to mission-orbit insertion.}
\label{fig:TESSwholemission}
\end{figure}

A graphic depicting the entire P/2-HEO mission, from launch to mission-orbit insertion, appears in Fig.~\ref{fig:TESSwholemission}.
Whether using supplemental propulsion (e.g., a solid rocket motor) or direct injection via the launch vehicle into an initial ``phasing'' orbit as low as 100,000 km, the spacecraft's on-board propulsion raises the phasing-orbit apogee close to the altitude of the Moon's orbit. Operationally, apogee may be raised through multiple burns, as shown in Fig.~\ref{fig:TESSwholemission}. TESS orbit analysis incorporated 3.5 phasing orbits with nominal apogees of 250,000, 328,000, and 400,000 km. The phasing-orbit perigee altitude is assumed to be 600 km, but a lower initial perigee could be raised with a small maneuver at first apogee. The final phasing orbit has perigee in LEO, apogee at approximately 400,000 km radius, and a period of $\sim$11 days. The second apogee can be adjusted to maintain the time of flight in the phasing orbits ($\sim$29 days for TESS)  to ensure the proper encounter with the Moon at final apogee. For this analysis, the equatorial inclination of the phasing orbits is assumed to be 28.5 deg. Before the last phasing orbit, a small burn is performed at perigee to target the conditions of the lunar flyby.

The lunar flyby delivers the spacecraft into a ``transfer'' orbit (blue in Fig.~\ref{fig:TESSwholemission}) with apogee greater than the Moon's orbit and perigee raised above the GEO belt (6.6 $R_E$). The geometry of the transfer orbit is critical to ensuring that the final mission orbital parameters can be achieved. Following the flyby, the spacecraft continues outbound to Post Lunar Encounter Apogee (PLEA). It is possible for the spacecraft to proceed inbound immediately following the flyby, but subsequent analysis shows that the inbound options are less favorable in $\Delta V$. From PLEA, the spacecraft descends to Post Lunar Encounter Perigee (PLEP). At PLEP, the spacecraft performs its last maneuver, the Period Adjust Maneuver (PAM). PAM lowers apogee so that the orbit period is reduced to 13.7 days, in 2:1 resonance with the Moon. Following PAM, the spacecraft is in the mission orbit (the black orbit labeled ``P/2-HEO'' in Fig.~\ref{fig:TESSwholemission}). Figure~\ref{fig:TESSwholemission} provides a two-dimensional projection of the orbits into the lunar-orbit plane. In practice, all of these orbits (phasing, transfer, and mission) have a non-zero inclination with respect to the lunar plane and have some component coming into or out of the page. 

Orbital perturbations from Earth oblateness, the Sun, and the Moon could in principle make the mission orbit unstable, leading to a violation of mission constraints. The mission-orbit perigee may drop below the GEO belt after a few years, or the inclination may drop low enough to create mission-ending eclipses. The long-term stability of the mission orbit depends on its initial orbit elements, which are driven in turn by the transfer orbit. The mission orbit's initial perigee is the same as PLEP, and the mission orbit's initial inclination, argument of perigee, and RAAN match those of the transfer orbit. Ensuring the long-term stability of a mission orbit therefore demands finding an appropriate transfer orbit and a judicious selection of flyby conditions. 

The first phase of this analysis considers the transition from the transfer orbit to the mission orbit. The transfer orbit begins immediately following the lunar flyby, so the constraints imposed by the geometry of the lunar orbit will affect what transfer orbits are available. When the transfer orbit has been characterized geometrically, the Moon's gravity is introduced analytically into the lunar flyby to reduce the space of possible transfer orbits and to demonstrate its impact on the size of the PAM.

The TESS P/2-HEO mission orbit must satisfy two requirements to ensure a feasible mission: 1) all eclipses must be less than 6 hours, and 2) perigee must remain between 7 and 22 $R_E$ throughout the 4-year mission. Before a specific orbit solution is sought out, it is necessary to first independently examine the impact each of the constraints has on the orbital elements.

\section{Eclipses}

A robust mission orbit would ideally satisfy the 6-hour eclipse-duration constraint even under the worst-case orbit geometry. If the mission-orbit perigee and apogee are assumed fixed, then the maximum eclipse durations are driven largely by the ecliptic RAAN, argument of perigee, and inclination. Under a best-case condition, the inclination is non-zero and high enough so that many orientations of the orbit have no eclipses at all. Consequently, the analysis begins assuming that the ecliptic inclination is greater than $\sim$10 deg and eclipses can occur only at one of two possible locations in the orbit when the spacecraft crosses the ecliptic plane. The longest eclipses occur when the orbit's apogee is aligned with the shadow of the Earth. 

\begin{figure}[htpb]
\centering
\includegraphics[width=4in]{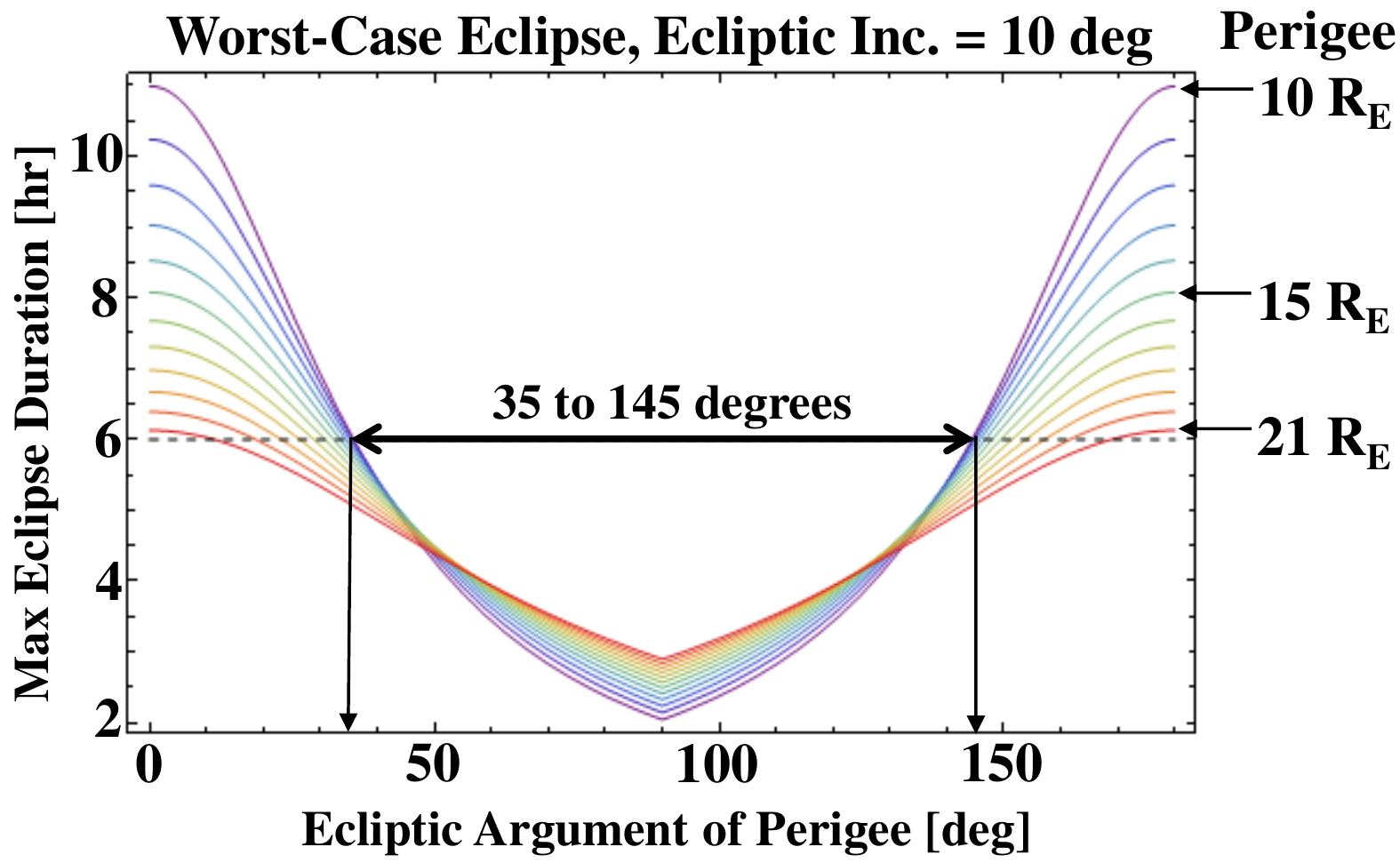}
\caption{Plot of worst-case eclipse times as a function of ecliptic argument of perigee for various mission-orbit perigees. Orbits with these RAANs and arguments of perigee may or may not be reachable; they are plotted here because they have the worst eclipse times of all possible mission orbits. The longest worst-case eclipses occur at apogee and can exceed 10 hrs depending on mission-orbit perigee. The shortest worst-case eclipses occur at an argument of perigee of 90 deg.}
\label{fig:eclipsesWorstCase}
\end{figure}

The worst-case eclipses occur for specific values of RAAN in the ecliptic plane: when the orbit's line of nodes is collinear with the Earth-Sun line. Under this condition, it is possible to cycle through ecliptic argument of perigee to evaluate all of the worst-case eclipses. Only some of these orientations have apogee in Earth's shadow, but they do represent worst possible eclipse orientations. Orbital ecliptic inclination at or above 10 deg does not have a significant impact on eclipse duration at the eccentricities under consideration. The worst-case set of eclipses appears in Fig.~\ref{fig:eclipsesWorstCase} for a range of mission-orbit perigees.
The sharp turn in the curve at the center of Fig.~\ref{fig:eclipsesWorstCase} corresponds to a switch in RAAN, where the node in shadow switches from ascending to descending.  The longest worst-case eclipses occur at apogee and can exceed 10 hrs depending on mission-orbit perigee. The shortest worst-case eclipses occur at an argument of perigee of 90 deg. Orbits with these RAANs and arguments of perigee may or may not be reachable, but they do represent the worst possible eclipse durations. The final mission orbit, for a given mission perigee, will have eclipses equal to or shorter than these eclipse durations. From the plots in Fig.~\ref{fig:eclipsesWorstCase}, satisfying the 6-hour eclipse constraint requires the mission-orbit perigee always to be above 21 $R_E$ or ecliptic argument of perigee to be between approximately 35 and 145 deg. Because the P/2-HEO perigee can vary below 21 $R_E$, a restriction on the argument of perigee is necessary. Some mission-orbit orientations (namely, those away from the worst-case RAANs) would permit lower or higher arguments of perigee and still satisfy the constraint, but to \textit{guarantee} that the eclipsing constraint is satisfied, the most conservative (and the selected) course of action is to restrict the orbit's ecliptic argument of perigee to the range of 35--145 deg.

\section{Minimum and Maximum Perigee}
Although the resonant motion of the P/2-HEO with respect to the Moon cancels much of the Moon's perturbing influence, the mission orbit is still subject to lunisolar perturbations that cause variations in the orbital elements. Under many circumstances it is challenging to predict how orbital elements are affected by third-body perturbations, but in the case of the P/2-HEO, generalized perturbation theory---namely, that of Kozai~\cite{Kozai:1962}---can help predict the mission orbit's qualitative behavior.

Kozai demonstrated that, for a small third-body perturbation, above a critical inclination (approximately 23 deg) the argument of perigee librates around 90 deg, as shown in Fig.~\ref{fig:TESSorbels} for the TESS DRM. More precisely, the inclination and argument of perigee form a closed loop (i.e., are periodic) in phase space. Furthermore, Kozai identified an approximately conserved quantity $K$ when the orbit is subject to the third-body perturbation:
\begin{equation}
K = \cos i \sqrt{1-e^2} = \cos i \sqrt{(r_p/a)(2- r_p/a)}\; ,
\label{eq:Kozai}
\end{equation}
where $i$ is the orbital inclination, $e$ the eccentricity, $r_p$ the periapsis, and $a$ the semimajor axis. This conserved quantity implies a tradeoff between inclination and eccentricity. As the orbit moves forward in time, the behavior of the orbit in phase space suggests that the inclination will increase or decrease, which must therefore be accompanied by a change in eccentricity. Because the P/2-HEO orbit period is in resonance with the Moon and nearly constant, the variation in eccentricity corresponds to a change in perigee, which threatens to violate the requirement that perigee remain between 7 and 22 $R_E$ for the entire mission.

\begin{figure}[h]
\centering
\includegraphics[width=4in]{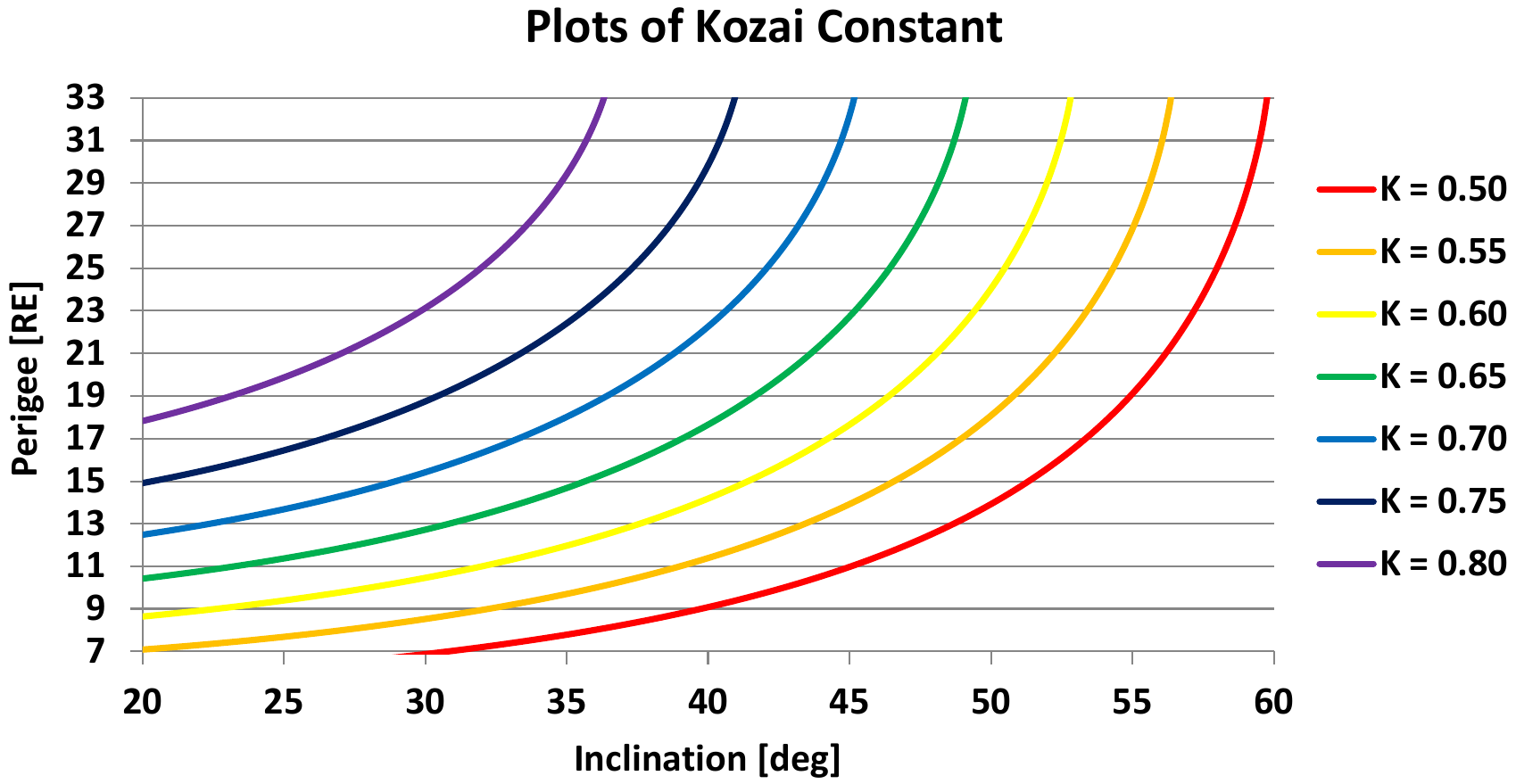}
\caption{P/2-HEO perigee as a function of inclination for several values of Kozai constant $K$. To satisfy the requirement that perigee remain between 7 and 22 $R_E$ for the entire mission, this plot shows that $K$ should be between approximately 0.60 and 0.70, with 0.65 appearing to be ideal.}
\label{fig:Kozai}
\end{figure}

Figure~\ref{fig:Kozai} shows a plot of the P/2-HEO perigee as a function of its lunar inclination for several values of Kozai constant $K$. For the P/2-HEO, $a$ in Eq.~\ref{eq:Kozai} is approximately 37.9 $R_E$. The maximum perigee is limited to this value of $a$, in which case the maximum possible inclination in Fig.~\ref{fig:Kozai} is $\arccos K$.
A value of $K$ for the TESS mission orbit was selected so that no matter how the inclination increases or decreases, the perigee remains within the range 7--22 $R_E$ over the mission life. The plots in Fig.~\ref{fig:Kozai} suggest that $K$ should be between 0.60 and 0.70, with 0.65 appearing to be ideal. Any lower than 0.60 and perigee runs the risk of dipping into the GEO belt; any higher than 0.70, and perigee may grow beyond 22 $R_E$. Over the decade-long oscillation period of the orbital elements, Fig.~\ref{fig:Kozai} shows that even a $K$ of 0.65 may exceed the 22 $R_E$ upper limit when the inclination grows large. However, the oscillation period is long enough that during the 4-year mission duration, no perigee constraint is violated. A violation does occur in the 25-year propagation in Fig.~\ref{fig:TESSorbels} around the year 2025, 8 years after the start of the propagation. Given that Kozai's theory does not provide a perfect model of the orbit's behavior,  a value of $K = 0.65$ offers margin on both the low and high ends of the desired perigee range. Equation~\ref{eq:Kozai} also indicates that once PLEP is chosen, $K$ is uniquely defined by the transfer-orbit inclination. To achieve the desired variability for the P/2-HEO, an initial inclination and PLEP can be targeted with Fig.~\ref{fig:Kozai} for the transfer and mission orbits that yield the Kozai constant $K$. Then the mission-orbit perigee and inclination will vary in accordance with Eq.~\ref{eq:Kozai} and Fig.~\ref{fig:TESSorbels}.

The analysis performed thus far, even before a specific orbit has been selected, has already narrowed down the trade space for the P/2-HEO. By looking at worst-case eclipse scenarios, the orbit's ecliptic argument of perigee should be between 35 and 145 deg to ensure eclipses less than 6 hours. Application of Kozai's perturbation theory shows that a Kozai constant $K = 0.65$ yields the most desirable behavior to maintain perigee between 7 and 22 $R_E$ throughout the mission. Furthermore, Kozai's observation about the libration of the argument of perigee suggests that the natural perturbed behavior of the orbit will move the initial argument of perigee ($>$35 deg) up towards 90 deg (about which it librates), which is the most desirable orientation to ensure short or no eclipses. As Eq.~\ref{eq:Kozai} suggests, the lunar inclination following the flyby sets the transfer- and initial mission-orbit inclination and impacts the variability of orbit perigee over the mission life.

\section{Finding Geometrically Feasible Orbits}
Thus far, no mention has been made about the feasibility for the desired orbital elements of the P/2-HEO. Desirable ranges of mission-orbit perigee, inclination, and argument of perigee have been identified, but can they be reached? Based on the mission concept previously described, it is clear that the mission orbit inherits its orbital elements from the post-flyby transfer orbit. The mission-orbit and transfer-orbit inclinations and arguments of perigee are the same, and the initial mission-orbit perigee is equivalent to PLEP. Therefore, successully achieving the desired orbital elements in the post-flyby transfer orbit by definition passes them on to the mission orbit. Consequently, the next several sections focus on the geometry of the lunar-flyby phase of the mission and how different flyby conditions affect the transfer orbit.

\begin{figure}[t]
\centering
\includegraphics[width=2.5in]{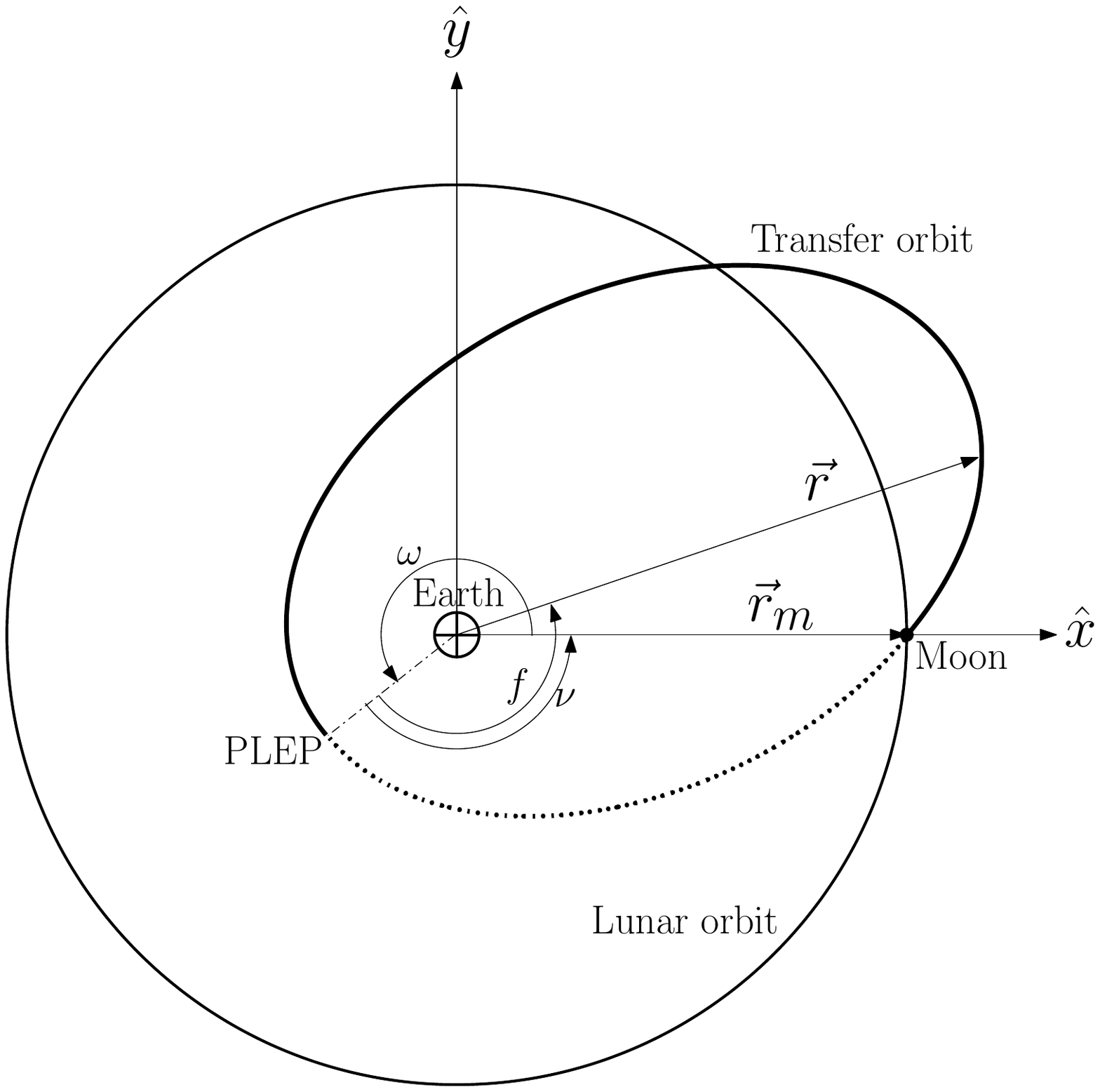}
\caption{The Earth-Moon geometry at the time of the flyby, using the Moon's orbital plane as the fundamental plane for calculation. The transfer orbit has a non-zero inclination, so some of the motion is into or out of the page.}
\label{fig:lunarPlane}
\end{figure}

From the several orbital planes relevant to the problem (i.e., equatorial, lunar, and ecliptic), the analysis of the lunar flyby and its associated constraints uses the Moon's orbital plane as the fundamental plane. It is first assumed the Moon follows a circular orbit about the Earth at a radius equal to the Moon's semimajor axis (this constraint is relaxed later). At the time of the flyby, the Earth-Moon geometry is frozen so that the $x$-axis is defined along the Earth-Moon line, and the $y$-axis is perpendicular to $\hat{x}$ in the direction of the Moon's orbital motion, as depicted in Fig.~\ref{fig:lunarPlane}. The $z$-axis completes the right-handed set, extending out of the page.
The assumption of a circular lunar orbit permits writing the position vector of the Moon as a function of time,
\begin{equation}
\vec{r}_m(t) = r_m\left[(\cos n_m t) \hat{x} + (\sin n_m t) \hat{y}\right] \; ,
\label{eq:MoonPosition}
\end{equation}
where $r_m$ is the radius of the Moon's orbit, $n_m$ is the Moon's mean motion, and $t$ is the time measured since the flyby. The position of the spacecraft is given in general as a function of its true anomaly $f$,
\begin{align}
\vec{r}(f) = &r(f)\left[\left(\cos\Omega\cos\theta-\sin\Omega\cos i\sin\theta\right)\hat{x}\right. \nonumber \\
&+ \left.\left(\sin\Omega\cos\theta + \cos\Omega\cos i\sin\theta\right)\hat{y} + \left(\sin i\sin\theta\right)\hat{z}  \right] \; ,
\end{align}
where $\theta \equiv \omega + f$, and the right ascension of the ascending node, $\Omega$ (measured from the $x$-axis), the argument of perigee, $\omega$, and the inclination, $i$, are measured with respect to the \textit{lunar orbit plane} using the coordinates shown in Fig.~\ref{fig:lunarPlane}. The radius of the spacecraft's position, $r(f)$, is given by the standard conic equation $r(f) = p/(1+e\cos f)$, where $p$ is the semilatus rectum and $e$ the orbital eccentricity. The Moon's perturbing influence on the spacecraft orbit is ignored, akin to the patched-conic approach for interplanetary mission design~\cite{Prussing:1993}.

At the moment of the flyby, $t = 0$. Therefore,
\begin{equation}
\vec{r}_m(t = 0) = r_m \hat{x}
\end{equation}
at the flyby. The true anomaly of the spacecraft immediately after the flyby (that is, the true anomaly on the transfer orbit at the moment of the flyby) is defined as a number $\nu$. To ensure that the spacecraft position coincides with the Moon requires
\begin{equation}
\vec{r}_m(t = 0) = \vec{r}(f = \nu) \; .
\label{eq:flybyCondition}
\end{equation}
On a component-by-component basis, Eq.~\ref{eq:flybyCondition} implies that
\begin{equation}
\sin i\sin(\omega+\nu) = 0 \quad\Rightarrow\quad \omega = -\nu \textrm{ or } \pi-\nu
\end{equation}
and 
\begin{equation}
\sin\Omega\underbrace{\cos(\omega+\nu)}_{=\pm 1}+\cos\Omega\cos i \underbrace{\sin(\omega+\nu)}_{=0} = 0 \quad\Rightarrow\quad \Omega = 0 \textrm{ or } \pi \; .
\end{equation}
The third vector component (not yet considering the vector magnitudes) provides an additional constraint on $\omega$ and $\Omega$,
\begin{equation}
\cos\Omega\cos(\omega+\nu)-\underbrace{\sin\Omega}_{=0}\cos i\underbrace{\sin(\omega+\nu)}_{=0} = 1 \quad\Rightarrow\quad \cos\Omega\cos(\omega+\nu) = 1 \; ,
\end{equation}
which implies that
\begin{align}
\Omega = 0 &\quad \textrm{when}\quad \omega = -\nu \label{eq:lunarConstraint1} \\
\Omega = \pi &\quad\textrm{when}\quad \omega = \pi-\nu \label{eq:lunarConstraint2} \; .
\end{align}
The requirement that $\Omega = 0$ or $\pi$ makes geometrical sense because the flyby is defined to occur on the $x$-axis in the lunar orbit plane, which means that the node (ascending or descending) must occur there as well. 

The Moon and spacecraft radii relative to the Earth must also be equal at the time of the flyby. If the radius of the spacecraft orbit is given by
\begin{equation}
r(f) = r_p(1+e)/(1+e\cos f) \; ,
\end{equation}
where $r_p$ is the radius of perigee, then this constraint requires
\begin{equation}
r(f=\nu) = r_m \quad\Rightarrow\quad \cos\nu = (1/e)[(r_p/r_m)(1+e)-1] \; .
\label{eq:lunarConstraint3}
\end{equation}
Equations~\ref{eq:lunarConstraint1},~\ref{eq:lunarConstraint2}, and~\ref{eq:lunarConstraint3} are three constraints imposed on the transfer orbit by the geometry of the flyby. That is, for the lunar flyby to occur, the orbital elements of the transfer orbit must satisfy those three equations.

\begin{figure}[t]
\centering
\includegraphics[width=4in]{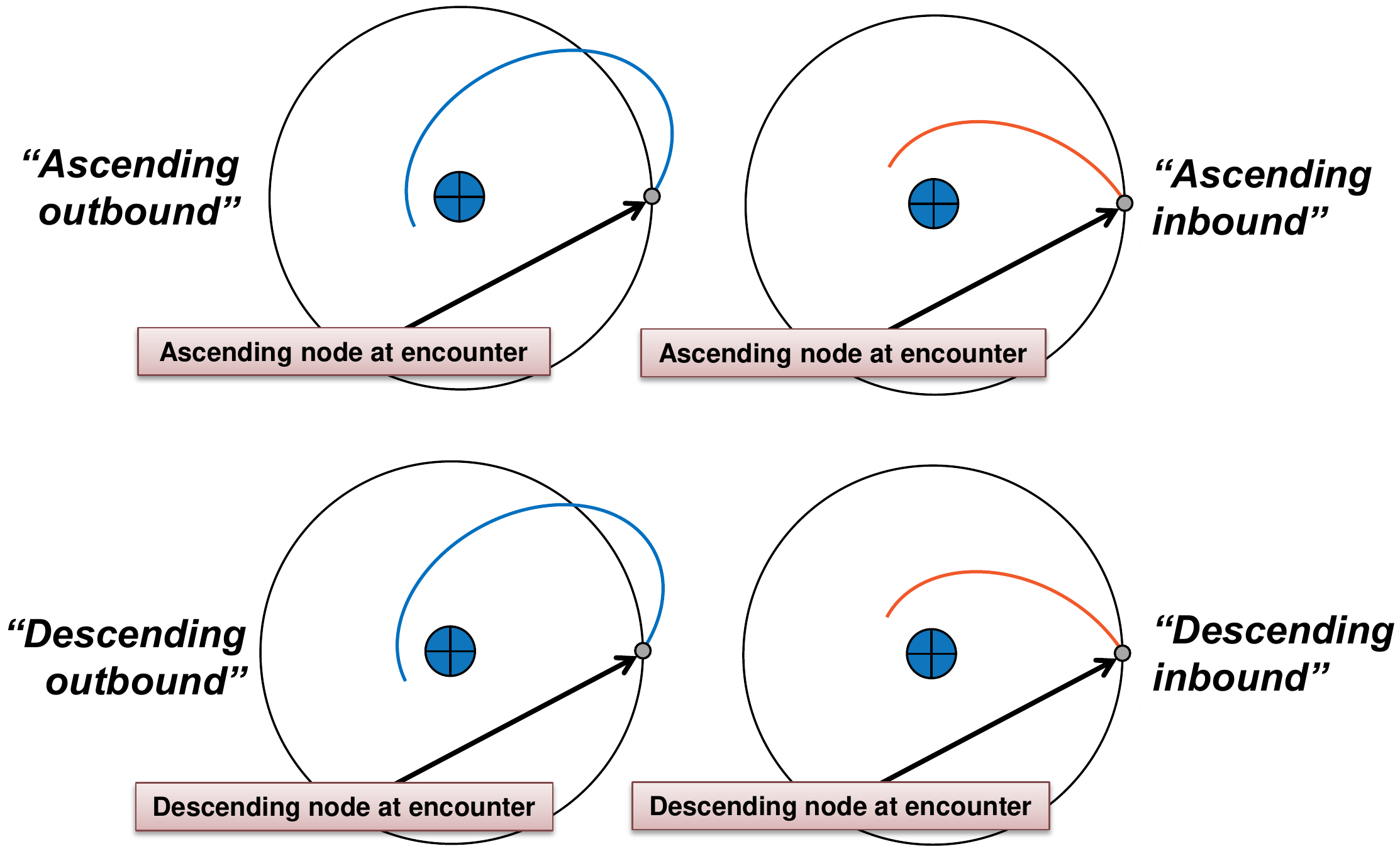}
\caption{The four possible paths of the spacecraft following the lunar flyby.}
\label{fig:transferOptions}
\end{figure}

Following the flyby, the spacecraft can continue on four possible paths, depicted in Fig.~\ref{fig:transferOptions}.
The spacecraft can either continue ``outbound'' towards apogee or make a sharp turn ``inbound'' back to the Earth, and the transfer orbit may depart the Moon either with an ascending node (out of the page) or a descending node (into the page). These four options have an impact on both the total PAM $\Delta V$ and the final argument of perigee of the transfer and mission orbits. The following analysis identifies which of these options is most desirable.

Regardless of the path the spacecraft takes on the transfer orbit, the P/2-HEO requires that the mission-orbit apogee be offset by 90 deg with respect to the Moon, so that lunar perturbations at apogee roughly cancel each month. This requirement implies that the Earth-spacecraft-Moon angle is 180 deg, as shown in Fig.~\ref{fig:TESSwholemission}. McGiffin~\cite{McGiffin:2001} notes that the apogee alignment can vary by up to 30 deg and yield a stable P/2-HEO, but the present analysis considers an exact alignment, to be relaxed later. The PLEP-alignment constraint limits the feasible space of orbital elements for the transfer orbit. To identify how the required PLEP alignment affects the transfer orbit, it is necessary to consider the relationship between the Moon's location at the post-PAM apogee (i.e., the first mission-orbit apogee) and the spacecraft's position at PLEP, which defines the line of apsides for the transfer and mission orbits and is directly related to the initial argument of perigee (Fig.~\ref{fig:TESSwholemission}). 

After the flyby, the spacecraft continues on the transfer orbit to PLEP in a time $T$ given by
\begin{equation}
T = P - M/n = (1/n)(2\pi - E + e\sin E) \; ,
\end{equation}
where $P$ is the transfer-orbit period, $n$ is the mean motion of the transfer orbit, $M$ is the mean anomaly of the flyby location on the transfer orbit, and $E$ is the eccentric anomaly of the flyby location on the transfer orbit. Using textbook transformations from eccentric to true anomaly, the time $T$ can be related to the true anomaly at the flyby, $\nu$ (cf.~Eq.~\ref{eq:lunarConstraint3}), by
\begin{equation}
T = \frac{1}{n}\left[2\pi - \arctan\left(\frac{\sqrt{1-e^2}\sin\nu}{e + \cos\nu} \right) + \frac{e\sqrt{1-e^2}\sin\nu}{1+e\cos\nu} \right] \; ,
\end{equation}
taking note that the inbound and outbound options for the transfer orbit may require a quadrant check for the mean anomaly. The spacecraft position unit vector at PLEP is
\begin{equation}
\hat{r}(PLEP) = (\cos\nu)\hat{x} - (\cos i\sin\nu)\hat{y} \mp (\sin i\sin\nu)\hat{z} \; ,
\end{equation}
where the $\mp$ option of the $z$-component corresponds to ascending versus descending options on the transfer orbit. Since alignment considerations with the Moon are strictly relative to the lunar plane, the projection of $\hat{r}$ in the lunar plane is defined,
\begin{equation}
\vec{r}'(PLEP) = \hat{r} - (\hat{r}\cdot\hat{z})\hat{z} = (\cos\nu)\hat{x} - (\cos i\sin\nu)\hat{y} \; .
\end{equation}

After PLEP and PAM, the spacecraft continues for another quarter period of the Moon's orbit, $(1/4)P_m$, to apogee, at which point the Moon's position \textit{in the lunar orbit plane} should be perpendicular to $\vec{r}'(PLEP)$:
\begin{equation}
\hat{r}_m[T + (1/4)P_m] \cdot \vec{r}'(PLEP) = 0 \; .
\label{eq:PLEPalignmentConstr}
\end{equation}
Performing this dot product and a few algebraic manipulations yields
\begin{equation}
\cos i = \cot\left[n_m\left(T+\frac{1}{4}P_m\right)\right]\cot\nu \; .
\label{eq:PLEPalignmentInc}
\end{equation}
Equation~\ref{eq:PLEPalignmentInc} restricts the lunar inclination of the transfer orbit and ensures that, following the lunar flyby, the subsequent PLEP is aligned with the Earth-Moon line. Equation~\ref{eq:lunarConstraint3} relates $\nu$, the transfer-orbit true anonaly at flyby, to the transfer orbit's eccentricity and PLEP. 

\section{Including the Moon's Gravity}
To include the effect of the Moon's gravity during a flyby, this analysis uses a patched-conic approximation with Tisserand's Criterion, which is an analytical expression that relates the orbital elements of the spacecraft before and after the lunar flyby~\cite{Strange:2002}. Tisserand's Criterion is closely related to the ``Jacobi Constant'' in multi-body dynamics and states that a constant $C$ is preserved during a flyby, where $C$ is given by
\begin{equation}
C = \frac{r_m}{a} + 2\cos i \sqrt{\frac{a}{r_m}\left(1-e^2\right)} = \frac{2r_m}{r_a + r_p} + 2\cos i \sqrt{\frac{2 r_a r_p}{r_m(r_a+r_p)}} \; ,
\label{eq:Tisserand}
\end{equation}
where $r_m$ is included because the Moon is the flyby body, $a$, $e$, and $i$ are the orbital elements before or after the flyby, $r_a$ and $r_p$ are the apogee and perigee radii, and $i$ is measured relative to the \textit{lunar} plane. The Tisserand constant of the post-flyby transfer orbit, $C_{\textrm{\tiny trans}}$, depends on only two variables, as $\cos i$ can be replaced by Eq.~\ref{eq:PLEPalignmentInc}. 

The pre-flyby Tisserand constant comes from the final phasing orbit, $C_{\textrm{\tiny phas}}$. The perigee and apogee for $C_{\textrm{\tiny phas}}$ are known based on the selection of a final phasing orbit with a perigee altitude of approximately 600 km and apogee radius of $\sim$400,000 km. In the limit that apogee is much greater than perigee (as is the case for the phasing orbits), Eq.~\ref{eq:Tisserand} can be approximated by
\begin{equation}
C_{\textrm{\tiny phas}} \approx 2r_m/r_a + 2\cos i \sqrt{2r_p/r_m} \; .
\label{eq:TisserandApprox}
\end{equation}
The phasing-orbit inclination is known in the \textit{equatorial} plane to be 28.5 deg, as delivered from a launch at Cape Canaveral Air Force Station (CCAFS), but $C_{\textrm{\tiny phas}}$ depends on the inclination in the \textit{lunar} plane. The transformation from equatorial to lunar inclination ranges by 38 deg depending on the date of the transformation at lunar flyby. The 28.5 deg inclination in the equatorial plane can vary between about 8.5 and 46.5 deg in the lunar plane. Therefore, for a fixed apogee and perigee of the final phasing orbit, the value of $C_{\textrm{\tiny phas}}$ ranges from a maximum when the lunar inclination of the phasing orbit  is 8.5 deg to a minimum when the lunar inclination is 46.5 deg.

\begin{figure}
\centering
\subfigure[Constant lunar radius.]{\includegraphics[width=2.5in]{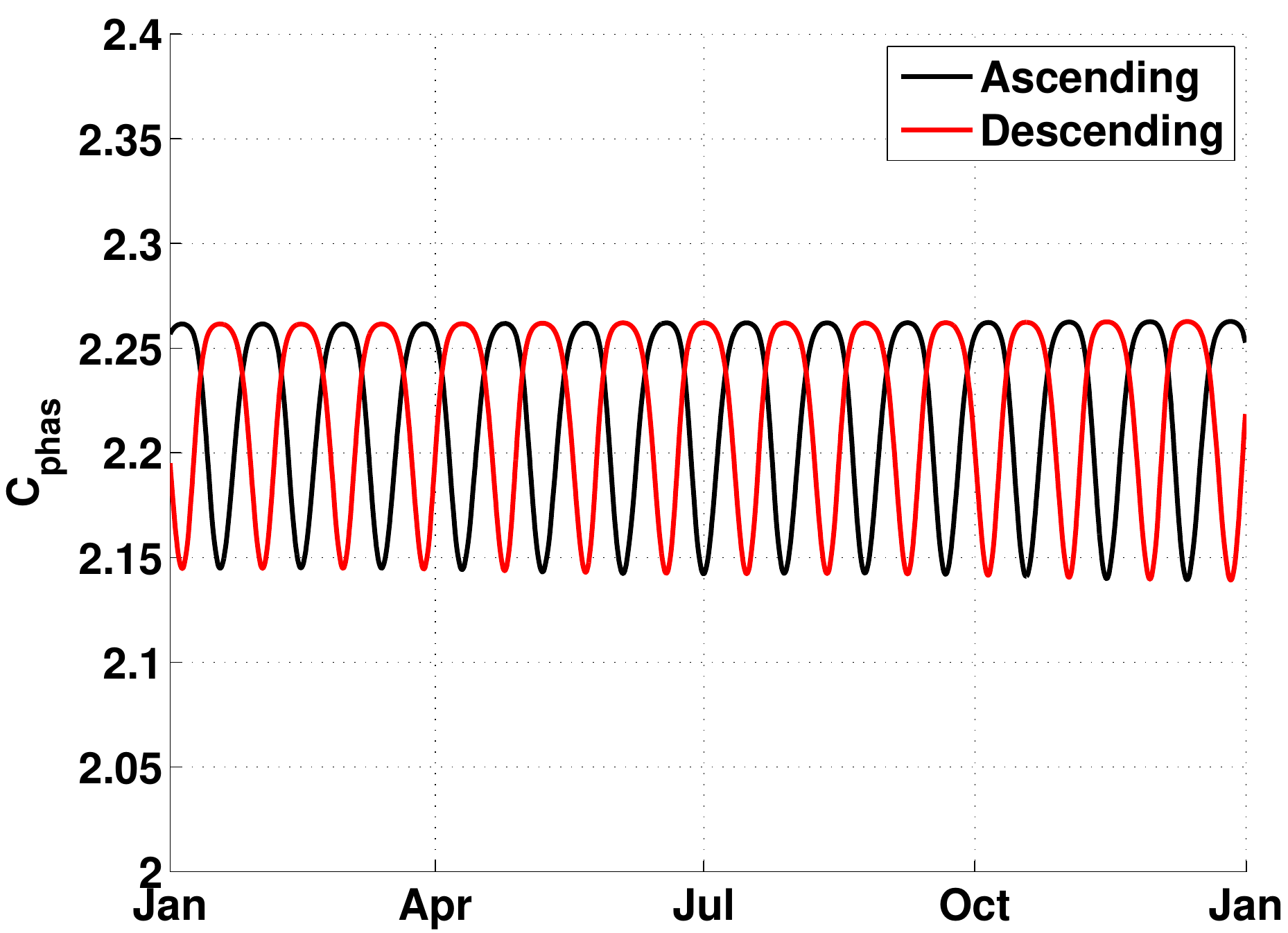}\label{fig:CphasConstRm}}\qquad
\subfigure[Time-varying lunar radius.]{\includegraphics[width=2.5in]{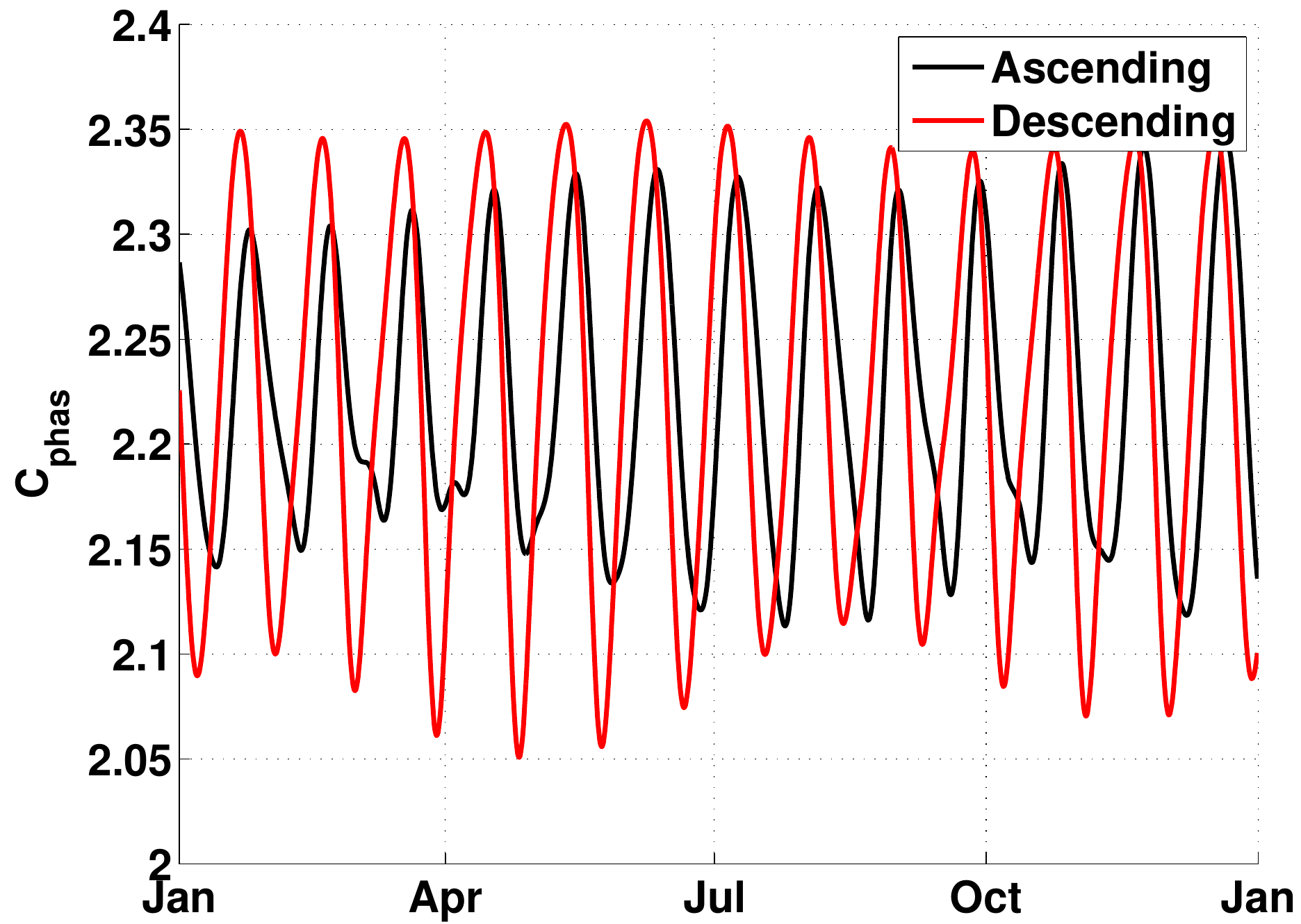}\label{fig:CphasTimeVaryRm}}
\caption{The Tisserand constant $C_{\textrm{\tiny phas}}$ for a 600 km altitude $\times$ 400,000 km radius phasing orbit with launch from CCAFS. The equatorial inclination of the phasing orbit is 28.5 deg, but the lunar inclination changes throughout the year, affecting $C_{\textrm{\tiny phas}}$. The plot at left shows $C_{\textrm{\tiny phas}}$ if the Moon radius $r_m$ is assumed fixed throughout the year. The plot at right shows the $C_{\textrm{\tiny phas}}$ if the Moon's true, time-varying radius is used.}
\label{fig:CphasYear}
\end{figure}

A plot of the time-dependence of $C_{\textrm{\tiny phas}}$ for a 600 km altitude $\times$ 400,000 km phasing orbit in the year 2017 appears in Fig.~\ref{fig:CphasYear}. At left in Fig.~\ref{fig:CphasConstRm} is a plot of $C_{\textrm{\tiny phas}}$ if the radius of the Moon's orbit, $r_m$, is assumed constant throughout the year, as is assumed throughout the foregoing analytical treatment. At right in Fig.~\ref{fig:CphasTimeVaryRm} is a plot of the same $C_{\textrm{\tiny phas}}$ transformation using the Moon's true, time-varying radius in 2017.
For a constant lunar-orbit radius (the Moon's semimajor axis), $C_{\textrm{\tiny phas}}$ varies from 2.14 to 2.25 about a mean value of 2.21. For the year 2017, the Moon's radius varies from 381,000 to 406,600 km. This range has a significant impact on the maximum and minimum $C_{\textrm{\tiny phas}}$, which varies from 2.01 to 2.39. If a particular $C_{\textrm{\tiny phas}}$ is desirable, it is possible to manipulate the phasing orbit, at the expense of extra $\Delta V$, to achieve the desired value, by changing either apogee or perigee. Equation~\ref{eq:TisserandApprox} suggests that changing apogee has the largest impact on $C_{\textrm{\tiny phas}}$ in the limit that perigee is small. 

Although $C_{\textrm{\tiny phas}}$ is variable, the range of variability occurs on a monthly time-frame. Numerical analysis later in this paper focuses on the month of June 2017, and the plot in Fig.~\ref{fig:CphasYear} suggests that the conclusions are bounded and should apply to any month of the year. That is, the behavior of $C_{\textrm{\tiny phas}}$ does not vary wildly from month to month, so analysis that depends on $C_{\textrm{\tiny phas}}$ in one month should remain valid throughout the year with slight adjustment.

\begin{figure}
\centering
\includegraphics[width=4.5in]{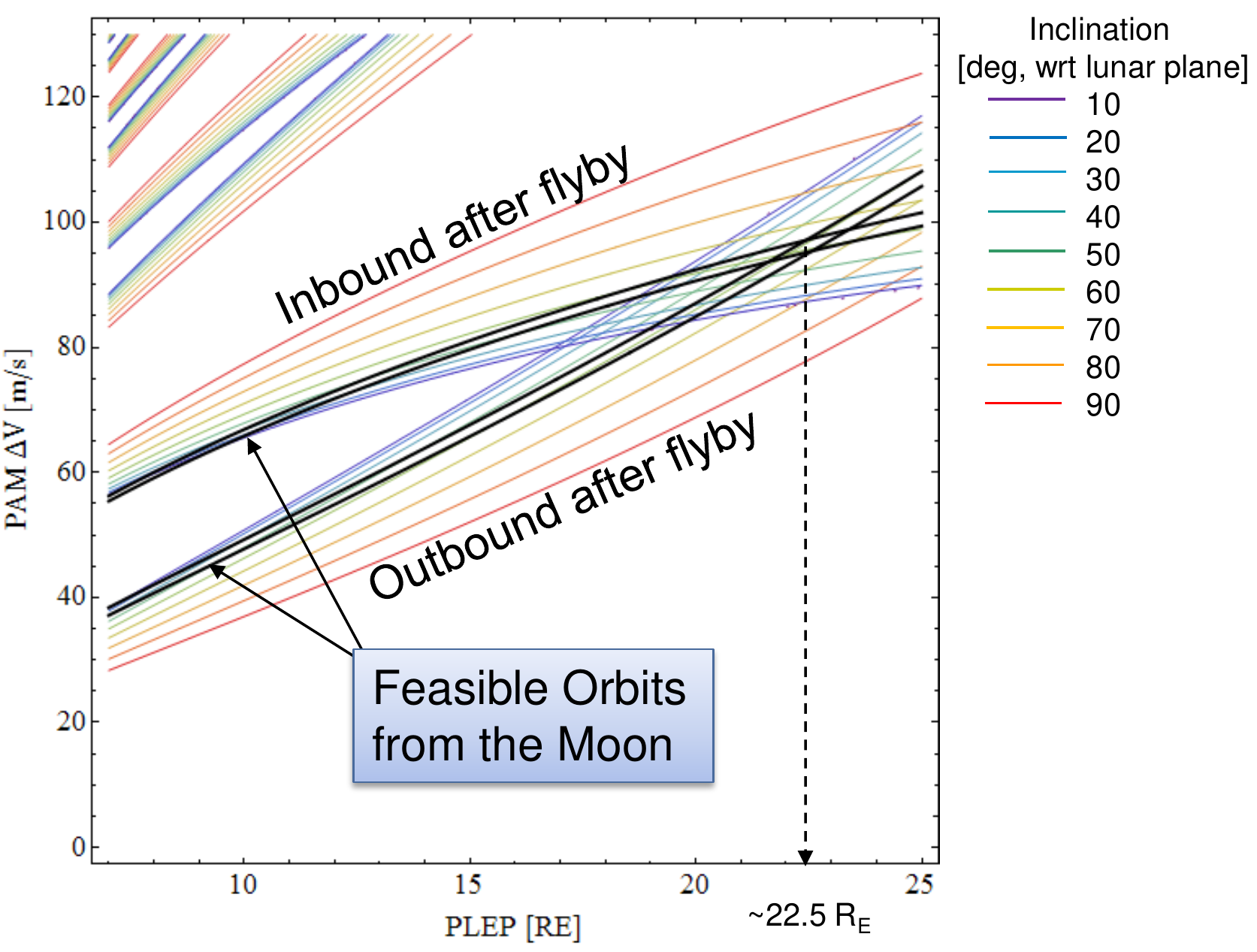}
\caption{Plots of post-flyby, transfer-orbit inclination as a function of PLEP and PAM $\Delta V$. The colored contours show transfer orbits that are \textit{geometrically} feasible by including the PLEP-alignment constraint. The \textit{gravitationally} feasible orbits are bounded between the pairs of superimposed black curves.}
\label{fig:rainbowTiss}
\end{figure}

Via textbook transformations, it is possible to translate the transfer-orbit eccentricity and PLEP into PAM $\Delta V$ because the post-PAM mission orbit period is known and PLEP is equivalent to the initial mission-orbit perigee. A plot of Eq.~\ref{eq:PLEPalignmentInc} as a function of PLEP and PAM $\Delta V$ appears in Fig.~\ref{fig:rainbowTiss}. The contours of constant transfer-orbit lunar inclination in Fig.~\ref{fig:rainbowTiss} appear in pairs, corresponding to inbound and outbound options. Integer multiples of the lunar orbit period (in Eq.~\ref{eq:PLEPalignmentInc}) yield additional pairs at higher $\Delta V$. 

Given the minimum and maximum $C_{\textrm{\tiny phas}}$, Figure~\ref{fig:rainbowTiss} shows plots of Eq.~\ref{eq:Tisserand} for the transfer orbit in black assuming a constant lunar-orbit radius. In other words, as Tisserand's Criterion demands $C_{\textrm{\tiny phas}} = C_{\textrm{\tiny trans}}$, the minimum or maximum $C_{\textrm{\tiny phas}}$ is inserted into Eq.~\ref{eq:Tisserand} and the resulting curve plotted as a function of PLEP and PAM $\Delta V$. If the time-varying lunar-orbit radius were used, the space between the pairs of black curves would widen marginally to encompass, for example, a year's worth of variability.
These black curves in Fig.~\ref{fig:rainbowTiss} are superimposed on the contours. The colored contours show the space of \textit{geometrically} feasible transfer orbits, and the space between the black curves denotes the subset that is \textit{gravitationally} feasible via a lunar flyby. Not all of the orbits in the subset are available at all times, because the equatorial-to-lunar inclination transformation for the phasing orbit depends on the date. The two pairs of black contours in Fig.~\ref{fig:rainbowTiss} cross at a PLEP of approximately 22.5 $R_E$. As the lower-$\Delta V$ pair of contours corresponds to the ``outbound'' transfer orbit and the mission requires PLEP to remain below 22 $R_E$, it is now possible to conclude that all feasible transfer orbits that subsequently meet mission-orbit constraints should follow the \textit{outbound} option.

\begin{figure}
\centering
\includegraphics[width=6in]{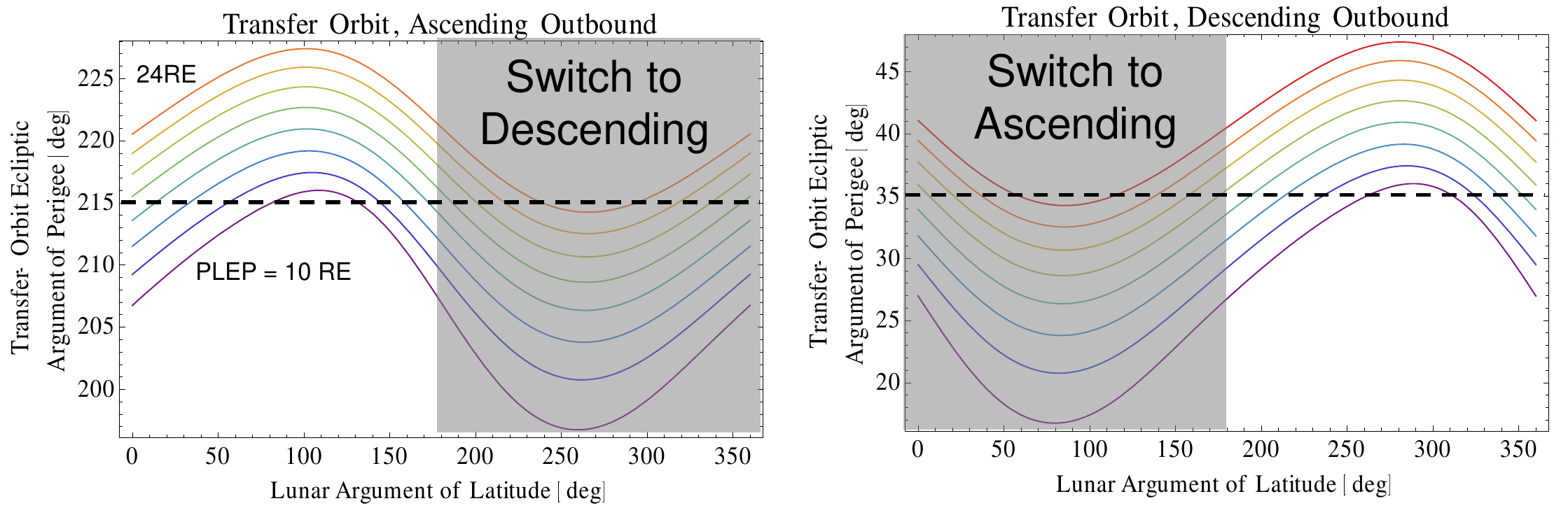}
\caption{The transfer-orbit ecliptic argument of perigee as a function of the lunar argument of latitude at the time of the flyby. To ensure that the apogee and perigee are more than 35 deg out of the ecliptic plane, the selection of ascending versus descending options switches depending on the lunar argument of latitude.}
\label{fig:Flip}
\end{figure}

It remains to be seen, however, whether the ascending or descending outbound option is preferable. For different values of PLEP, Eqs.~\ref{eq:lunarConstraint1} and~\ref{eq:lunarConstraint2} permit a direct calculation of the lunar argument of perigee of the transfer orbit through Eqs.~\ref{eq:lunarConstraint3} and~\ref{eq:PLEPalignmentInc} (also shown in Fig.~\ref{fig:TESSwholemission}). However, the eclipsing requirement on argument of perigee from Fig.~\ref{fig:eclipsesWorstCase} is based on the \textit{ecliptic} argument of perigee. The transformation between lunar and ecliptic argument of perigee depends not only on the orbital elements of the lunar orbit but also the position of the Moon in its orbit at the time of the flyby. The Moon's position is measured by the Moon's ecliptic argument of latitude (i.e., the Moon's angular distance from its ascending-node crossing with the ecliptic plane). By applying various rules of spherical trigonometry to the orbital elements of the Moon in the ecliptic and lunar planes, it is possible to plot the transfer orbit's \textit{ecliptic} argument of perigee as a function of lunar argument of latitude. The cases for ascending and descending outbound options appear in Fig.~\ref{fig:Flip}.

For eclipse purposes, the higher the ecliptic argument of perigee relative to the ecliptic plane, the better. For the ascending outbound case in Fig.~\ref{fig:Flip}, the ecliptic argument of perigee is between 215 and 225 deg (35--45 deg out of the ecliptic plane) early in the month for most PLEPs, but dips lower in the second-half of the month (i.e., when the lunar argument of latitude is greater than 180 deg). For descending outbound, the situation is reversed: early in the month, the ecliptic argument of perigee is lower compared to later in the month. Neither ascending nor descending is individually preferable. Instead, a hybrid solution is the key to ensuring high argument of perigee under all circumstances. Because ascending versus descending outbound remains a free choice for the mission designer, Fig.~\ref{fig:Flip} suggests that when the lunar flyby occurs in the first half of the month, the ascending outbound option should be selected for the transfer orbit. When the flyby happens in the second half of the month (i.e., after the Moon has crossed its descending node relative to the ecliptic), descending outbound should be selected. Under this scheme, the orbital elements of the transfer and mission orbits depend on the day of the month of the lunar flyby, with a discontinuity halfway through the month. This scheme guarantees an ecliptic argument of perigee at least 35 deg out of the ecliptic plane at the start of the mission for PLEPs greater than $\sim$16 $R_E$.

At the beginning of this analysis, the trade space for the P/2-HEO was large, involving the orbital elements of the phasing, transfer, and mission orbits, plus the time of the lunar flyby. The orbital elements of the phasing orbit are restricted by three requirements: 1) launch from CCAFS, which affects inclination, 2) perigee in LEO, and 3) apogee at the Moon to allow a lunar flyby. The phasing orbit's other elements are affected by the time of launch and the time of the lunar flyby. The orbital elements of the transfer orbit are restricted by four requirements: 1) minimal PAM $\Delta V$ requires the ``outbound'' option following the lunar flyby, 2) an ecliptic argument of perigee greater than 35 deg to minimize eclipse duration requires a switch between ``ascending'' and ``descending'' options depending on the Moon's location in its orbit at the time of the flyby, 3) the Earth-spacecraft-Moon alignment at PLEP is near 180 deg at PLEP, and 4) Tisserand's constant must be equal before the flyby ($C_{\textrm{\tiny phas}}$) and after ($C_{\textrm{\tiny trans}}$). The mission orbit is the most restricted of all: 1) by definition, the orbital period must be half of the Moon's, 2) the initial perigee is equal to the previously selected PLEP, and 3) the mission orbit and transfer orbit initial inclination, initial argument of perigee, and initial nodal longitude are the same.

Among all of these geometrical and dynamical constraints on the orbits, the only free variable is $C_{\textrm{\tiny phas}}$, which is equivalent to the day of the month of the lunar flyby, per Fig.~\ref{fig:CphasYear}. The $C_{\textrm{\tiny phas}}$ is a function of the phasing-orbit orbital elements, which may have been perturbed by the Sun and Moon (discussed in the next section). The phasing orbit delivers a $C_{\textrm{\tiny phas}}$ that, through the lunar flyby, yields an identical $C_{\textrm{\tiny trans}}$ for the transfer orbit. The mission requirements listed above, in conjunction with the goal of achieving a Kozai constant of $0.65 \pm 0.05$ for the mission orbit, fix all of the subsequent orbit parameters. If one knows the $C_{\textrm{\tiny phas}}$ and a desired $K$ (or, equivalently, transfer-orbit inclination per Eq.~\ref{eq:Kozai} and Fig.~\ref{fig:Kozai}), for a given PLEP, the rest of the mission sequence is determined.

\section{PLEP Misalignment}
\begin{figure}
\centering
\includegraphics[width=3.5in]{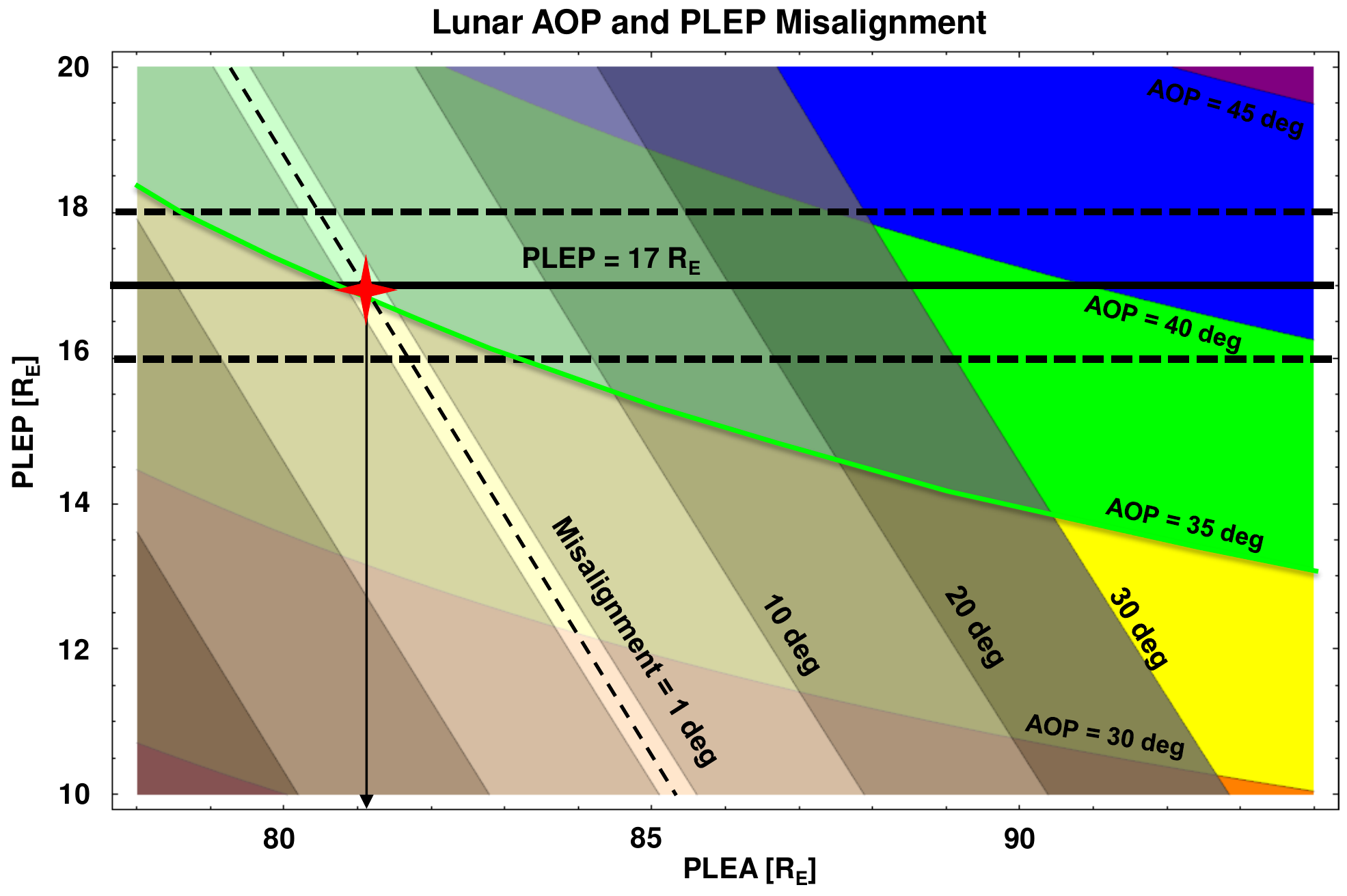}
\caption{Contours of lunar argument of perigee (AOP) and PLEP misalignment as functions of transfer-orbit PLEA and PLEP. The ideal case of AOP = 35 deg and 0 deg misalignment is marked by a red star, indicating a target PLEP of 17 $R_E$.}
\label{fig:PLEPselection}
\end{figure}

Although the desired Earth-spacecraft-Moon alignment at PLEP is 180 deg, the stable behavior of the P/2-HEO can tolerate a misalignment up to approximately 30 deg~\cite{McGiffin:2001}. Relaxing the PLEP-alignment constraint from a strict 0 deg to $\pm$30 deg introduces more freedom in the mission design. If one permits a PLEP misalignment as shown in Fig.~\ref{fig:TESSwholemission}, then the right-hand side of Eq.~\ref{eq:PLEPalignmentConstr} is non-zero, but a relationship between transfer-orbit inclination and PLEP akin to Eq.~\ref{eq:PLEPalignmentInc} is still available. Figure~\ref{fig:PLEPselection} shows a combination of contour plots that together provide the information necessary to select the most appropriate PLEP. The horizontal and vertical axes of Fig.~\ref{fig:PLEPselection} indicate the transfer-orbit PLEA and PLEP, respectively. PLEP has a strong correlation to argument of perigee, whereas PLEA more strongly influences the time of flight to PLEP and thus impacts the misalignment at PLEP. 

The colored contours show the lunar argument of perigee (AOP) of the transfer orbit (defined in Fig.~\ref{fig:TESSwholemission}), ranging from 45 deg in the upper right corner to 25 deg in the lower left. A green arc marks the desired minimum argument of perigee of 35 deg. Superimposed on these colored contours are greyscale contours of the PLEP misalignment. The ideal case of AOP = 35 deg and PLEP misalignment of 0 deg occurs when PLEP = 17 $R_E$ and PLEA = 82 $R_E$, as marked in Fig.~\ref{fig:PLEPselection} by a red star. For a PLEP of 17 $R_E$, the PLEA can range from 73 to 88 $R_E$ while keeping PLEP misalignment below 30 deg. This freedom in PLEA or PLEP misalignment is key to ensuring that one can find a mission orbit with a Kozai constant of $0.65 \pm 0.05$. For TESS, PLEP is selected to be 17 $R_E$ based on the requirement of a maximum mission-orbit perigee of 22 $R_E$. If a higher maximum mission perigee is allowed, PLEP can increase up to about 20 $R_E$, which allows higher initial AOP and yields shorter eclipse durations. However, increased PLEP requires additional $\Delta V$ for the PAM burn, as demonstrated in Fig.~\ref{fig:rainbowTiss}.

By relaxing the PLEP-alignment constraint, the choice of encounter date and PLEP does not completely determine the orbital elements of the mission orbit, but instead permits the choice of PLEA (and by proxy, lunar inclination), as long as the PLEP misalignment does not exceed $\pm$30 deg. Because the eccentricity of the mission orbit is fixed by perigee (PLEP) and the period (half the Moon's orbit period), varying the Kozai constant $K$ is equivalent to varying the lunar inclination. Using Eqs.~\ref{eq:Kozai} and~\ref{eq:Tisserand}, PLEA can be written in a simplified manner in terms of PLEP, $K$, and $C_{\textrm{\tiny phas}}$,
\begin{equation}
r_{\textrm{\tiny PLEA}} = \frac{2r_m}{C_{\textrm{\tiny phas}} - \eta K} - r_{\textrm{\tiny PLEP}} \; , \qquad \eta \approx 2^{2/3} \; ,
\label{eq:PLEAapprox}
\end{equation}
where $r_{\textrm{\tiny PLEA}}$ is the radius of PLEA and $r_{\textrm{\tiny PLEP}}$ the radius of PLEP. The value $\eta = 2^{2/3}$ is an approximate value derived through Taylor expansions with Eqs.~\ref{eq:Kozai} and~\ref{eq:Tisserand} by noting that PLEP is much smaller than PLEA and $r_m$ for most of the transfer orbits considered here. Equation~\ref{eq:PLEAapprox} illustrates that PLEA (as a proxy for semimajor axis of the transfer orbit) is the control variable to yield a $K$ value given a particular $C_{\textrm{\tiny phas}}$. Although Eq.~\ref{eq:PLEAapprox} provides guidance on the selection of PLEA, the PLEP misalignment must remain less than 30 deg.

\begin{figure}
\centering
\includegraphics[width=4in]{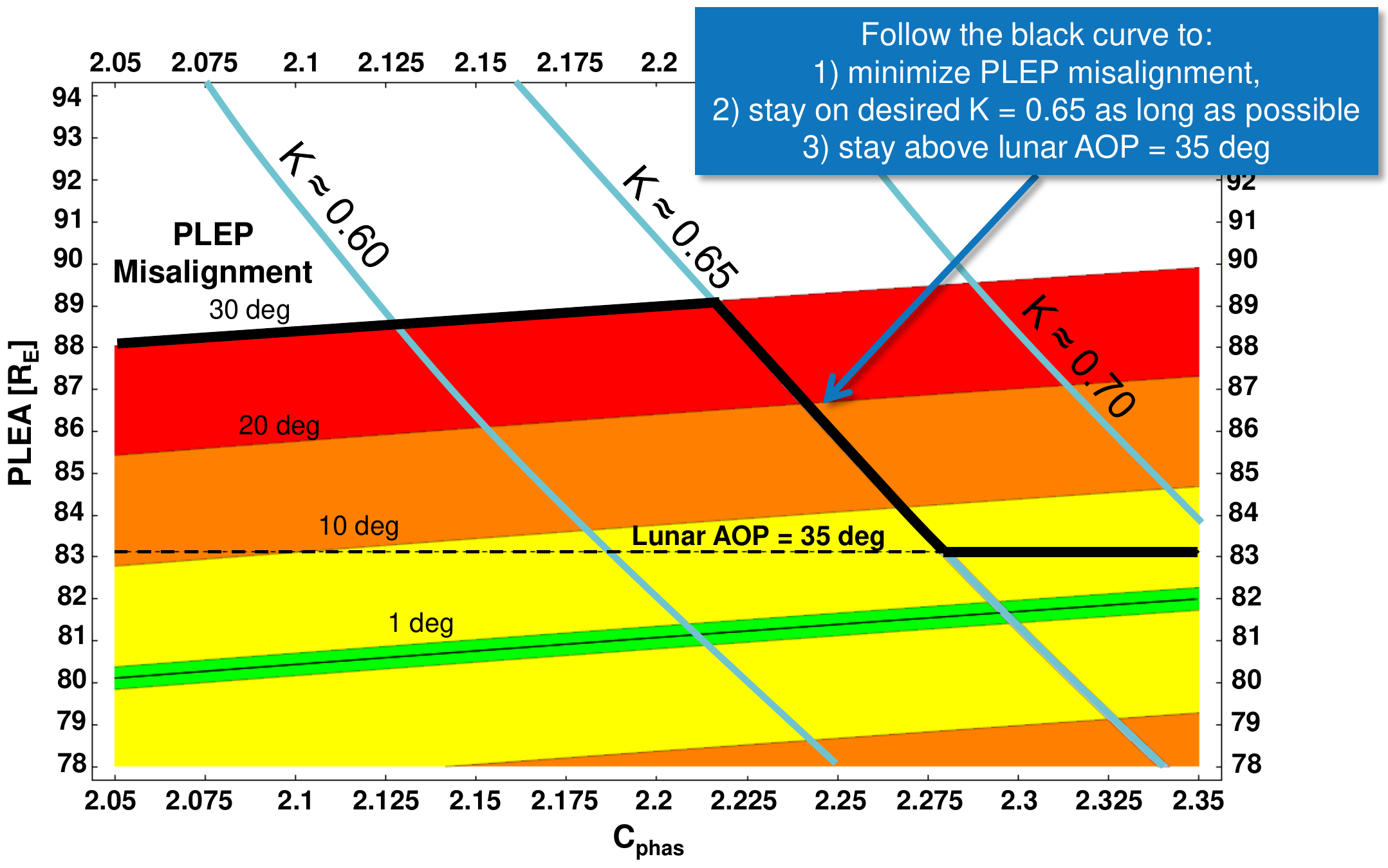}
\caption{A mission-design tool that combines several operational constraints to target transfer-orbit orbital elements. $C_{\textrm{\tiny phas}}$ is on the horizontal axis and PLEA is on the vertical axis. The colored contours represent the degrees of PLEP misalignment. Contours of different Kozai constants extend vertically, and a horizontal line is included where the lunar AOP = 35 deg. The black contour on the figure traces the mission design that most closely satisfies the constraints.}
\label{fig:PLEPMisalignment}
\end{figure}

Figure~\ref{fig:PLEPMisalignment} depicts a tool used to select (or target) the transfer-orbit orbital elements that lead to a desirable mission orbit. 
The figure shows the phasing-orbit Tisserand constant, $C_{\textrm{\tiny phas}}$, on the horizontal axis and the transfer-orbit PLEA on the vertical axis, with colored contours that represent the degrees of PLEP misalignment relative to the Moon's position (0 deg misalignment is most desirable). Also on this plot is a line where $K = 0.650$, the desired Kozai constant. A lunar argument of perigee (AOP) of 35 deg is indicated by the dashed line.  Given a $C_{\textrm{\tiny phas}}$ value, the orbit should satisfy the 30-deg misalignment constraint and have initial AOP $\ge$ 35 deg.  Then, by varying PLEA, the desirable Kozai constant of 0.65 can be realized. The black line on the figure traces the mission designs that most closely satisfy all these constraints. The PLEA in Fig.~\ref{fig:PLEPMisalignment} is the value targeted in high-fidelity simulations that, through the analytical guidance, should yield mission orbits that satisfy the mission's eclipse and perigee-range constraints. A PLEA below 83 $R_E$ can be used (above $C_{\textrm{\tiny phas}}$ = 2.3) for certain days of the lunar cycle when the transformation of lunar to ecliptic argument of perigee yields acceptable eclipse and alignment results. For $C_{\textrm{\tiny phas}}$ below 2.125, the desired $K$ range can no longer be achieved and minimum perigee may not remain above GEO.

\section{Phasing Orbits}
Regardless of spacecraft size, operational constraints, spacecraft propulsion system, or launch vehicle, the strategy for getting to the P/2-HEO orbit via lunar flyby starts with reaching a highly eccentric orbit with an initial apogee above 100,000 km and increasing to a final apogee of ~400,000 km that intersects the orbit of the moon. This could be done by direct injection, or through a series of phasing orbits that might also include a series of apogee-raise maneuvers. As an example, the TESS mission strategy shown in Figures~\ref{fig:TESSrot} and~\ref{fig:TESSwholemission}  uses a series of 3.5 phasing orbits with progressively increasing apogee, where the final apogee is fixed at the Moon's radius and a burn at the preceding perigee targets the desired lunar flyby conditions. The equatorial inclination of the phasing orbits is 28.5 deg, assuming launch from CCAFS, which drives the geometric variation of $C_{\textrm{\tiny phas}}$ previously discussed and presented in Figure~\ref{fig:CphasYear}. The time of flight (~29 days for TESS) allocated to the phasing orbits is also critical, and can be adjusted by the apogee of the second phasing orbit with minimal change in $\Delta V$.

In addition to the geometry-driven variation of $C_{\textrm{\tiny phas}}$, the phasing orbits are perturbed by other factors, such as lunar and solar gravity. These perturbations depend on both the relative positions of the Earth, Moon, and Sun and the duration of the phasing orbits. The higher the period of the phasing orbits and the greater the time spent at large apogee distances, the more the perturbations alter the final phasing orbit, thereby changing $C_{\textrm{\tiny phas}}$ and the subsequent transfer orbit.

In practice, the Tisserand constant evolves under the influence of perturbations during the transfer orbit from lunar flyby to PLEP as well. Tisserand's constant is based on the assumptions of the restricted three-body problem, so realistic trajectories diverge from this idealized case, especially for highly-eccentric, long-duration orbits such as the phasing and transfer orbits. Beyond this point in the analysis, high-fidelity tools are necessary to find fully integrated trajectories that include a full phasing-orbit sequence and targeted lunar flyby that achieve the desired initial conditions for the mission orbit.

\section{Mission-Orbit Variability and Eclipse Analysis}
The preceding analysis demonstrates that when the mission-orbit initial conditions are chosen carefully, the mission constraints can be satisfied for nearly any time during the lunar cycle. Given a launch epoch, eclipse constraint, and desired Kozai constant (bounded by the $\pm$30 deg lunar alignment requirement), the values of PLEP and PLEA, given $C_{\textrm{\tiny phas}}$, can be systematically chosen to achieve a desired mission orbit. To verify the analytical approach, the next step is to simulate the P/2-HEO orbit for the mission duration as well as over multiple decades as presented in Fig.~\ref{fig:TESSrot}. The P/2-HEO was initialized thousands of times through a grid of initial conditions and propagated at high fidelity. The simulations were initialized across the dimensions PLEP and lunar flyby epoch, but also $C_{\textrm{\tiny phas}}$ to test the sensitivity to the Tisserand constant, which is subject to vary from the analytical value under the influence of gravitational perturbations.

All propagations were performed with TRACE, a high-fidelity ephemeris-generation and orbit-determination program developed by The Aerospace Corporation. TRACE has a decades-long heritage at The Aerospace Corporation supporting the analysis and design of satellite orbits and tracking systems, including operational missions. The simulations presented here used an 8$\times$8 Earth-gravity model with third-body perturbations from a point-mass Sun and Moon.

The example analysis that follows was performed for the TESS trajectory design; consequently, the simulations use epochs during the lunar month that correspond to a candidate launch date for TESS. The range of lunar encounter dates in the simulation falls between the two lunar ascending nodes on 27 June and 25 July 2017, with the lunar descending node occurring on July 12. If an analyst were to apply the P/2-HEO to a different mission with a different mission date, the results would differ slightly by lunar cycle due to the changing relative position of the Sun, and $C_{\textrm{\tiny phas}}$ would vary from month-to-month as illustrated in Fig.~\ref{fig:CphasYear}. The initial conditions for the orbit trade space were studied for PLEPs from 7 $R_E$ (just above GEO) to 37 $R_E$ (circular orbit) in increments of 1 $R_E$, $C_{\textrm{\tiny phas}}$ from 2.0 to 2.4 in increments of 0.01, and over the entire lunar orbit in increments of 6 hours. Each set of initial conditions was propagated for 25 years, which was adequate to characterize the long-term evolution of the orbit and more than sufficient to characterize the four-year expected lifetime of the TESS mission.

\begin{figure}[t]
\centering
\subfigure[Minimum perigee.]{\includegraphics[width=2.75in]{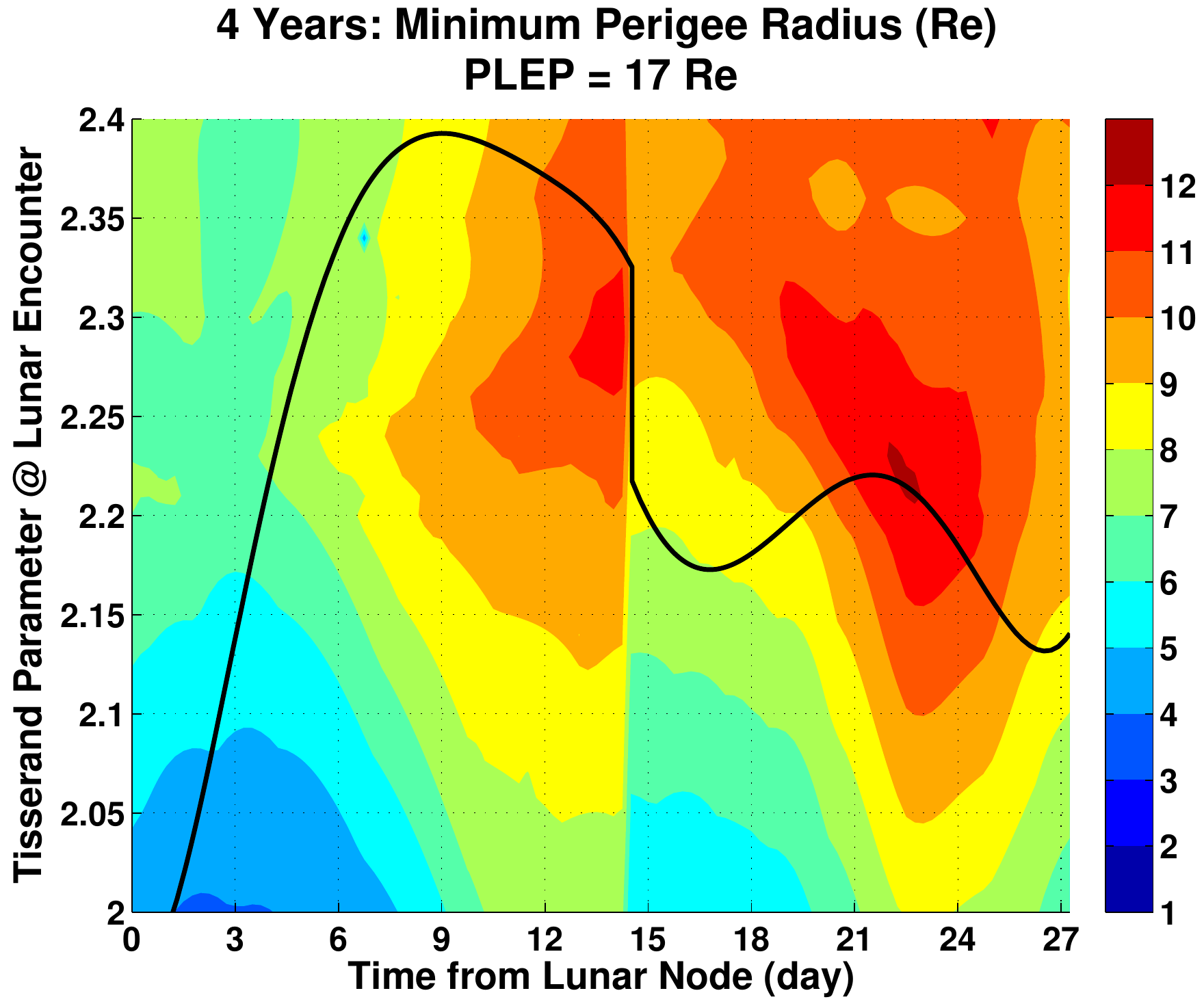}\label{fig:minPerigee}}\qquad
\subfigure[Maximum perigee]{\includegraphics[width=2.75in]{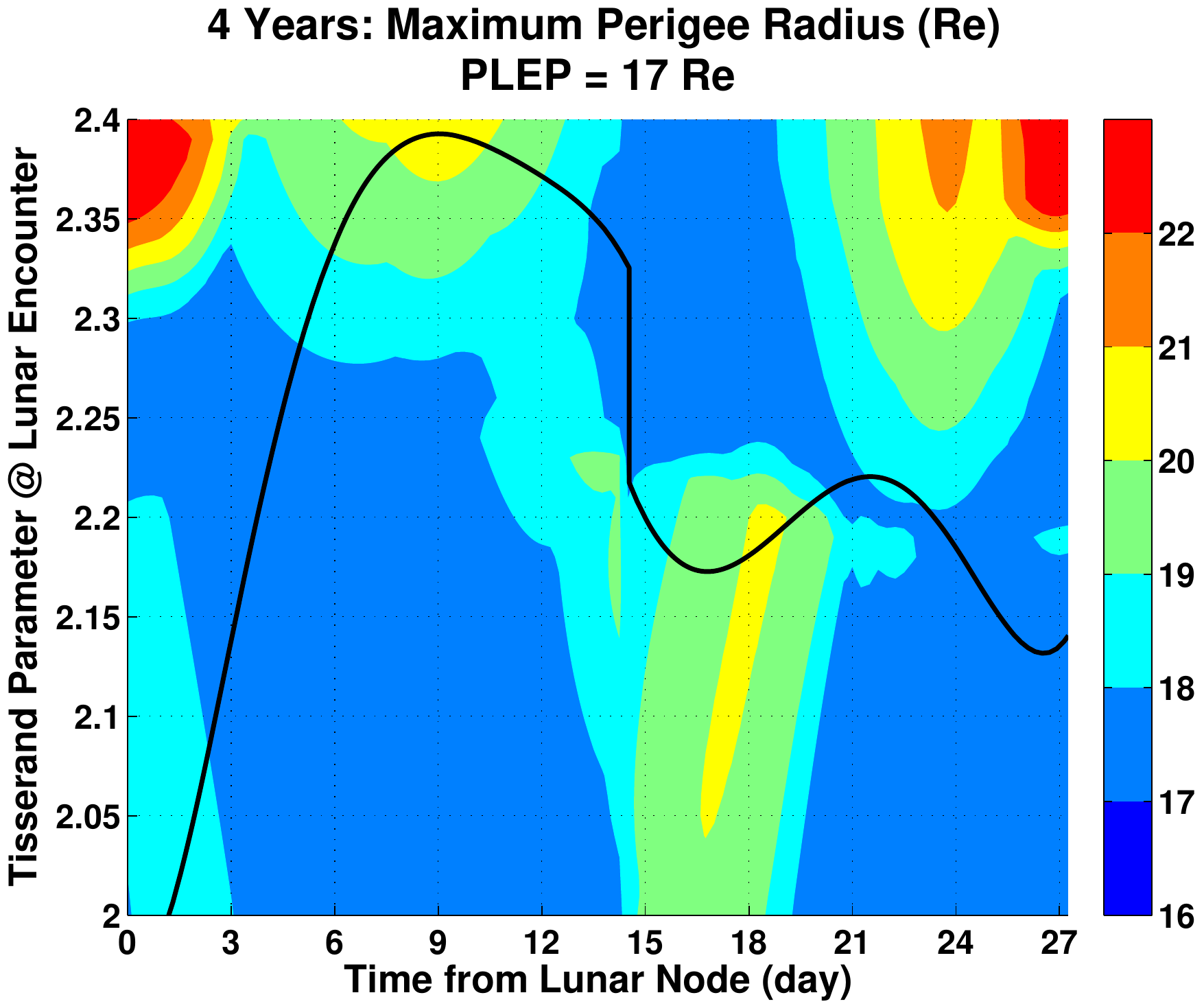}}
\caption{Contour plots of minimum and maximum perigee radius of the mission orbit over a 4-year propagation. The vertical axis is the initial Tisserand constant of the mission orbit; the horizontal axis is the time of the flyby relative to the lunar ascending node. The superimposed black line is the $C_{\textrm{\tiny phas}}$ available from the pre-flyby phasing orbit.}
\label{fig:ContourMinPerigee}
\end{figure}

\begin{figure}[t]
\centering
\includegraphics[width=3in]{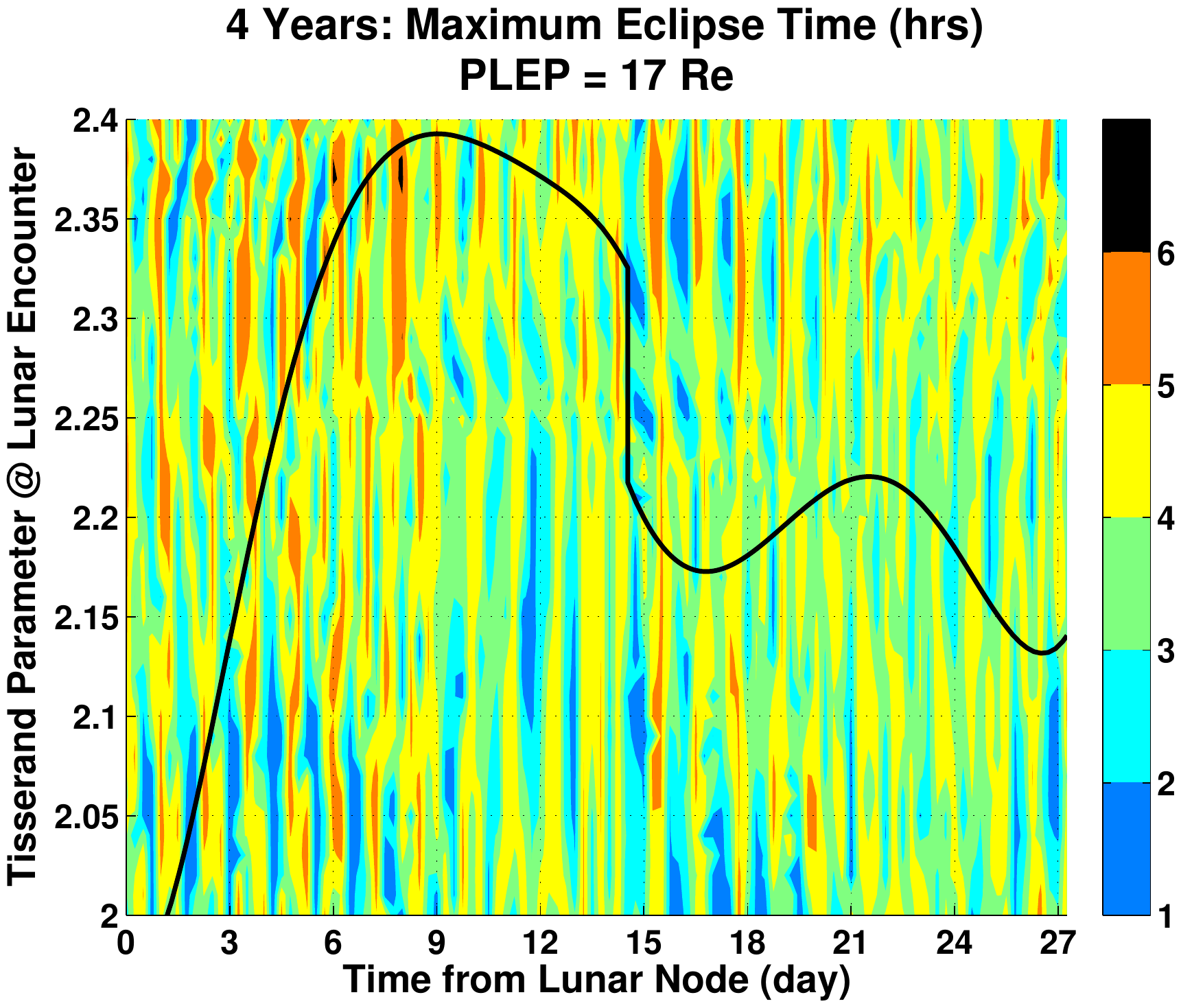}
\caption{Contour plots of maximum eclipse duration on the mission orbit over a 4-year propagation. The vertical axis is the initial Tisserand constant of the mission orbit; the horizontal axis is the time of the flyby relative to the lunar ascending node. The superimposed black line is the $C_{\textrm{\tiny phas}}$ available from the pre-flyby phasing orbit.}
\label{fig:ContourMaxEclipse}
\end{figure}

The plots in Figs.~\ref{fig:ContourMinPerigee} and~\ref{fig:ContourMaxEclipse} show contours of four-year simulation results for P/2-HEO missions with PLEP = 17 $R_E$. The time of the lunar encounter, measured from the Moon's ascending node, is measured on the horizontal axis and the Tisserand constant ($C_{\textrm{\tiny phas}}$) at the lunar encounter on the vertical axis. The color bar indicates the value of each parameter over the course of the simulation. The black line traces the path of attainable mission-orbit initial conditions given a particular phasing orbit approach to the lunar encounter (in this case, 3.5 TESS nominal phasing orbits) and includes all lunisolar perturbations. This line represents the $C_{\textrm{\tiny phas}}$ from the fully-patched trajectory, including all lunisolar perturbations from launch to flyby, with no aforementioned analytical approximations.

Figure~\ref{fig:ContourMinPerigee} shows the 4-year evolution of minimum and maximum perigee. Desirable mission orbits have a perigee that remains above the GEO belt (6.6 $R_E$) and below 22 $R_E$ for the entire mission. The areas in Fig.~\ref{fig:minPerigee} starting with the green contours (7 $R_E$) through red (12 $R_E$) yield these desirable orbits. The undesirable orbits reside in the lower-left corner of the plot (and repeat after the lunar descending node near day 14), early in the lunar cycle and where $C_{\textrm{\tiny phas}}$ is below ~2.1. The apparent discontinuity in the contours at day 14---the lunar descending node---occurs because of the decision earlier to switch from ascending- to descending-outbound transfer orbits depending on the lunar argument of latitude, a scheme that yields an ecliptic argument of perigee with desirable eclipse durations. Figure~\ref{fig:ContourMinPerigee} indicates that the design solution is robust to small variations in nominal values of $C_{\textrm{\tiny phas}}$ for the majority of the lunar cycle.  Lunisolar perturbations mainly have the effect of lowering and raising final phasing orbit perigee depending on the Sun's orientation during the lunar cycle, which alters $C_{\textrm{\tiny phas}}$ only slightly, as illustrated in Eq.~\ref{eq:TisserandApprox}. The monthly oscillation shown in Fig.~\ref{fig:CphasYear} has much more influence on the variation of $C_{\textrm{\tiny phas}}$.

Early in the analysis, it became clear that low PLEP values (7--10 $R_E$) resulted in multi-year perigee oscillations that dipped into the GEO region. At the high end of the PLEP range (25--37 $R_E$), the mission orbit is less eccentric, and the variability of the oscillations in the Earth-Moon three-body system began to break down, resulting in erratic changes in the apogee and perigee.  The stable region that satisfied mission constraints, as originally suggested by the above analysis with the Kozai mechanism, appeared to reside in between these two ranges. The choice of PLEP = 17 $R_E$ was made after confirmation of this detailed orbit analysis. Other choices of PLEP (e.g., 16--20 $R_E$ based on Fig.~\ref{fig:Kozai}) may be appropriate for missions with different applications and constraints.

Figure~\ref{fig:ContourMaxEclipse} shows the maximum eclipse duration for a four-year mission. This plot indicates that the eclipse behavior is more erratic than the perigee variability, where the difference between an orbit with a short or long eclipse is very sensitive to a small change in the encounter time. $C_{\textrm{\tiny phas}}$ has minimal impact on eclipse duration. Figure~\ref{fig:ContourMaxEclipse} validates that the choice of PLEP = 17 $R_E$ ensures the orbit is safe from eclipses of 6 hours or greater, as expected based on the earlier analysis (e.g., selecting initial conditions that ensure a high ecliptic argument of perigee). Although not discussed in this paper, eclipses in the phasing orbits can have durations greater than 6 hours for a few days each lunar cycle when the phasing orbit apogee is along the Earth-Sun line. This must be accounted for in the mission trajectory design.

\begin{figure}
\hspace*{\fill}
\subfigure[Minimum mission-orbit perigee.]{\includegraphics[width=2.4in]{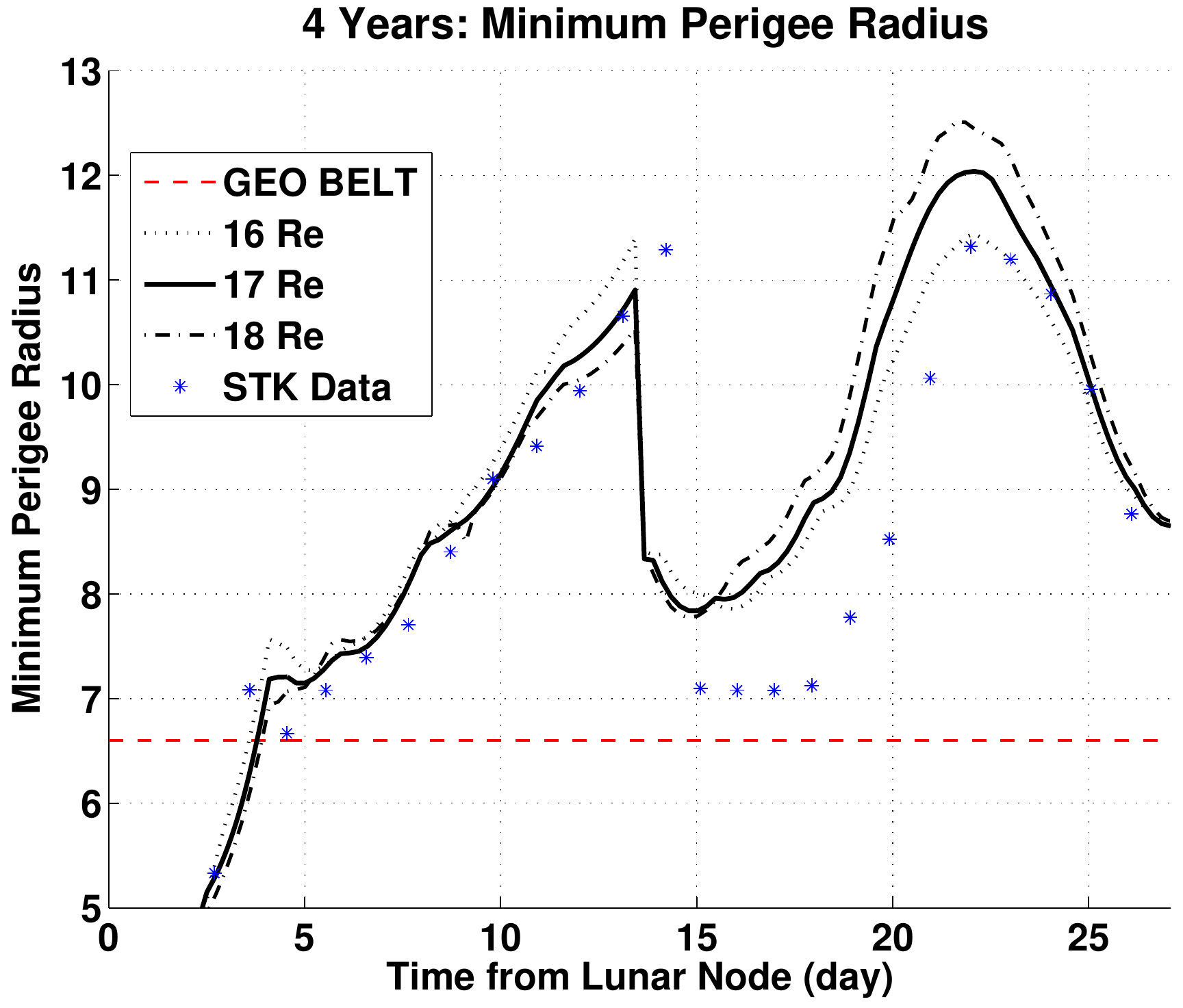}\label{fig:MinPerigeeSTK}}
\hfill
\subfigure[Minimum ecliptic inclination.]{\includegraphics[width=2.4in]{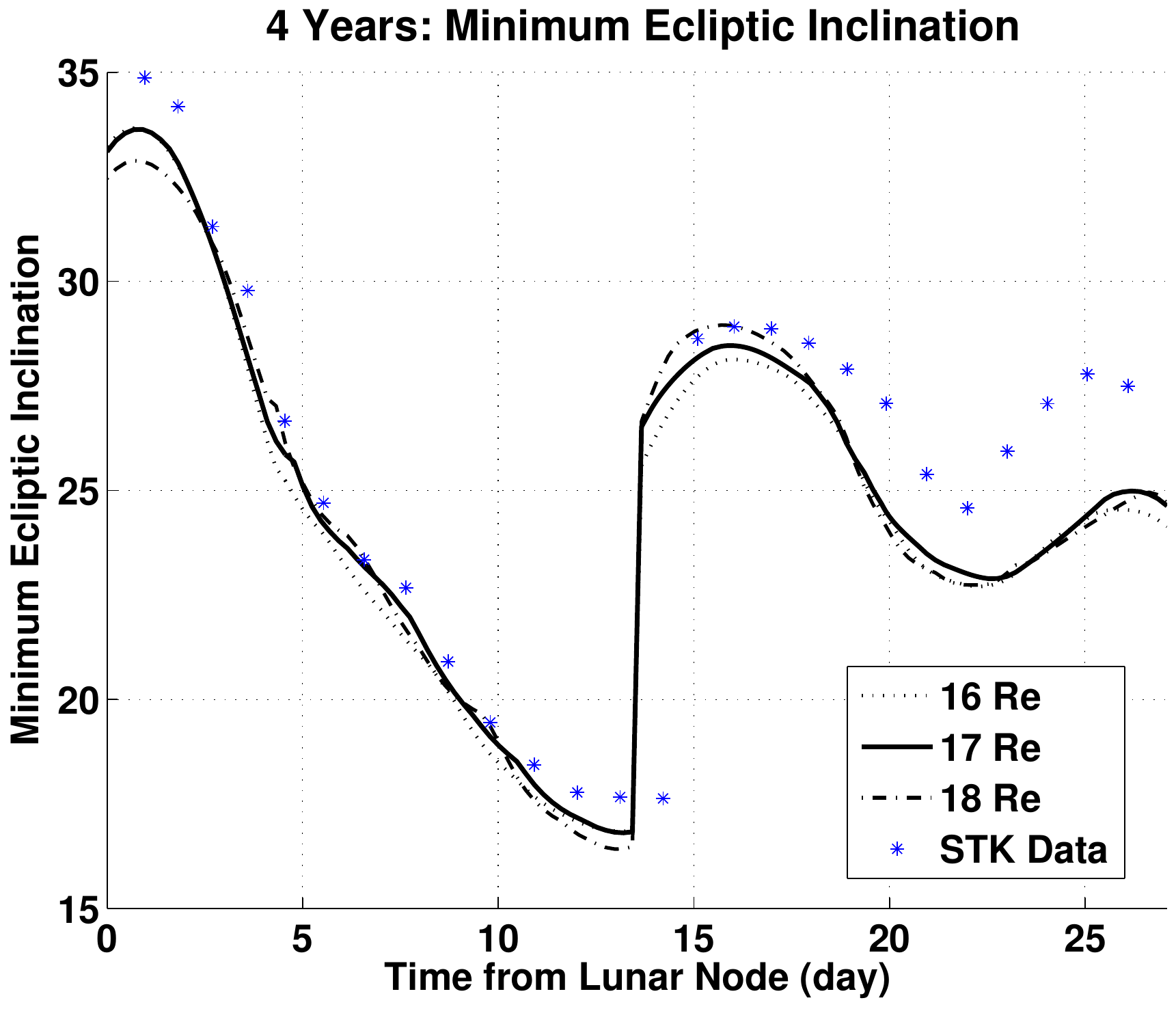}\label{fig:MinIncSTK}}
\hspace*{\fill}
\caption{A comparison of 4-year propagations with TRACE and STK. STK is fully patched; TRACE begins at the mission orbit.}
\label{fig:TRACEvsSTK}
\end{figure}

\begin{figure}[htb]
  \begin{minipage}[t]{0.483\linewidth}
    \centering
    \includegraphics[width=2.4in]{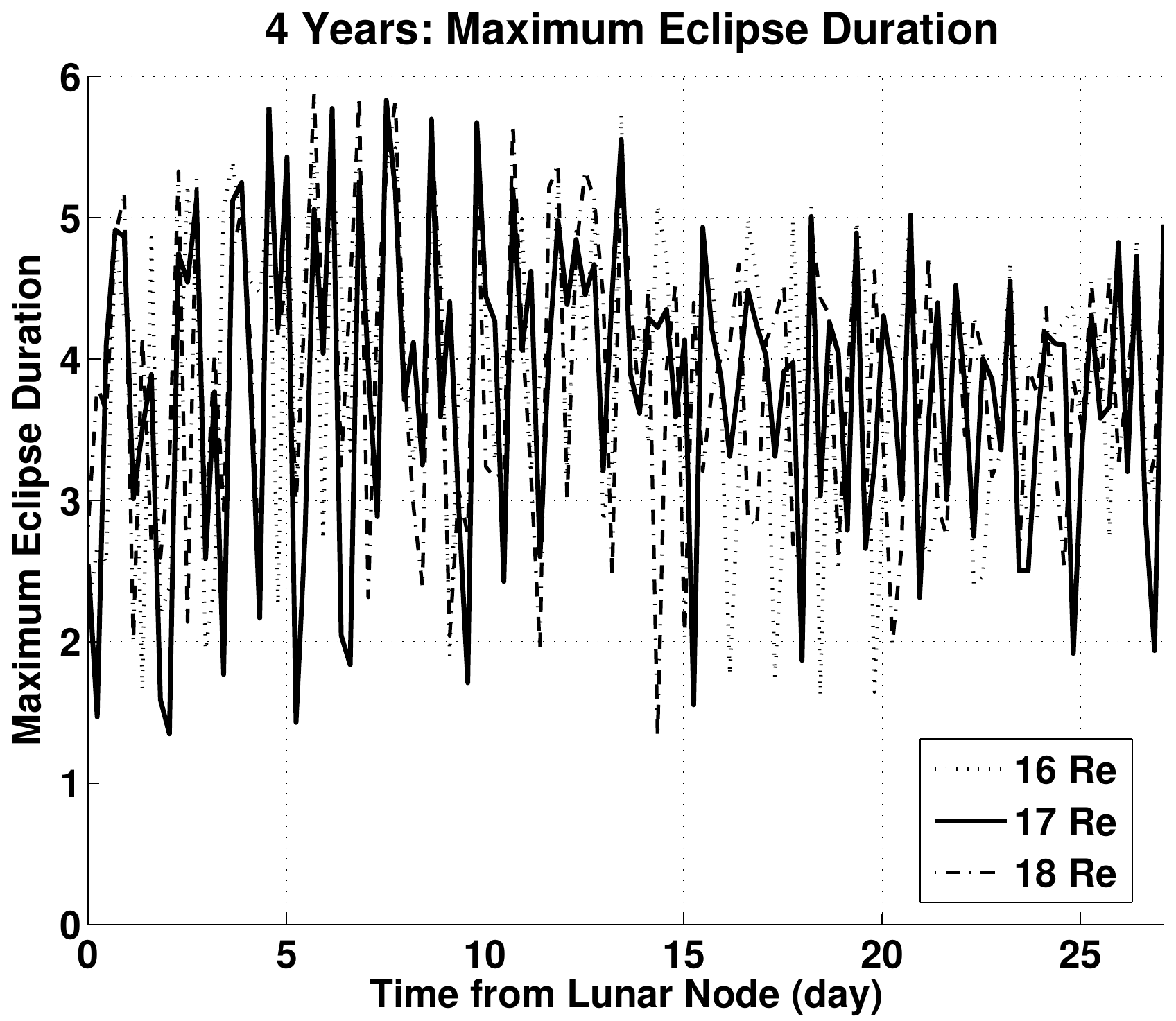}
    \caption{The 4-year maximum eclipse time evaluated by TRACE.}
    \label{fig:MaxEclipseTRACE}
  \end{minipage}%
  \hspace{0.5cm}
  \begin{minipage}[t]{0.483\linewidth}
    \centering
    \includegraphics[width=2.4in]{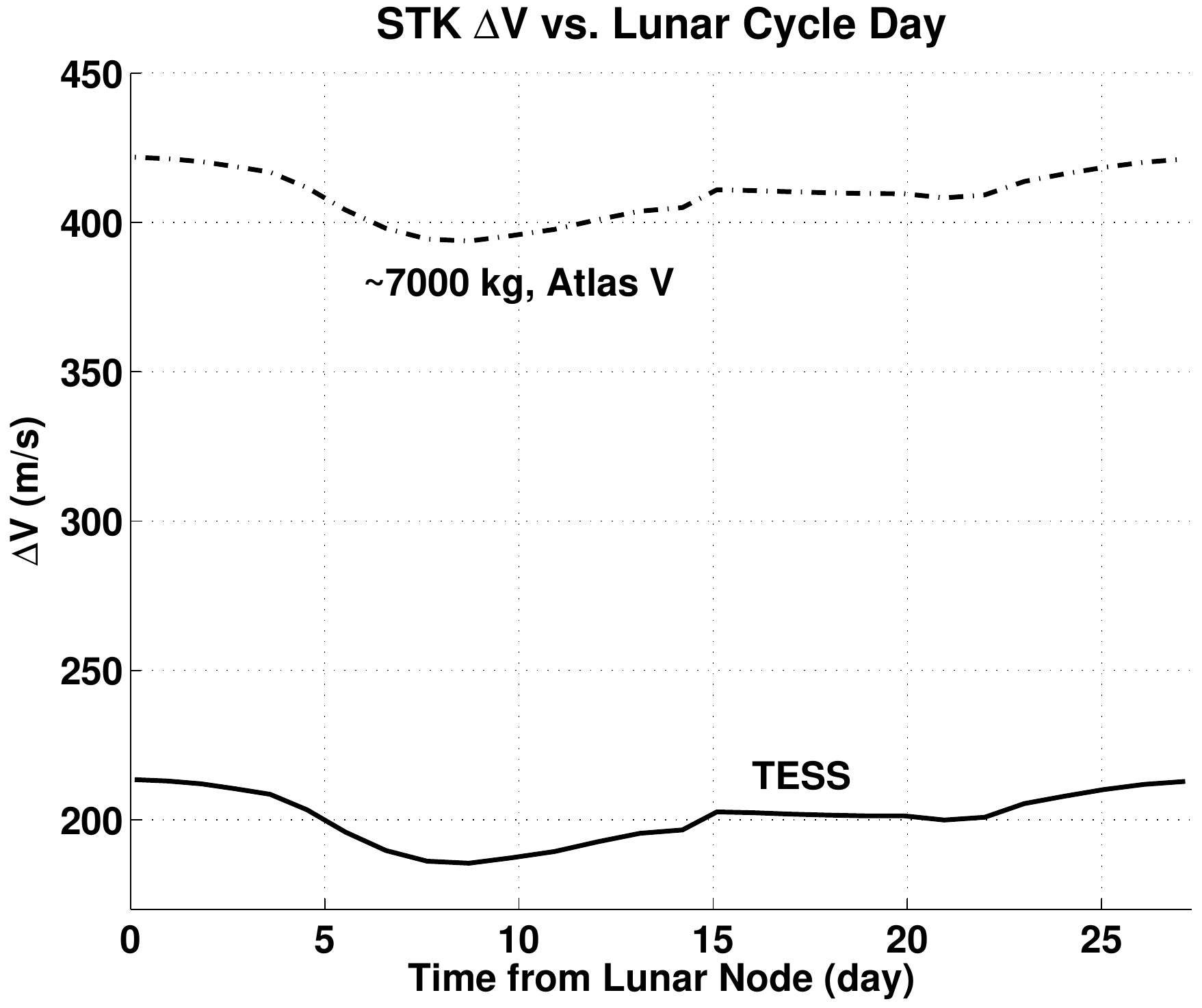}
    \caption{Total mission $\Delta V$---as evaluated in STK---as a function of day of the month of the lunar flyby.}
    \label{fig:STKdv}
  \end{minipage}
\end{figure}

To determine the mission-orbit performance that is actually achievable, only those orbits that lie along the black contour are available for more detailed analysis. TRACE is used again to simulate those orbits at 6-hour increments. The four-year minimum mission-orbit perigee, minimum ecliptic inclination, and maximum eclipse time for PLEPs ranging over 16--18 $R_E$ are shown in Figures~\ref{fig:MinPerigeeSTK}, \ref{fig:MinIncSTK}, and~\ref{fig:MaxEclipseTRACE}, respectively. Also plotted are targeted STK results that are discussed in the next section. Both the TRACE initial conditions and the STK trajectories were generated using the tool in Fig.~\ref{fig:PLEPMisalignment}. The main difference is that the STK trajectories begin at the initial phasing orbit, but the TRACE simulations begin at the mission orbit. The perigee-variation constraint is satisfied except for a few lunar-encounter days early in the lunar cycle.
Except for a few days of lunar encounter, the ecliptic inclination remains above 20 deg, which helps prevent long mission eclipses. The total number of eclipses does not exceed 12 for any of these orbits, and the longest eclipses always occur near the beginning of the mission or near the end, when the low ecliptic argument of perigee brings the apogee closest to the ecliptic plane. The 6-hour eclipse duration constraint is satisfied for the entire lunar cycle, although there are many trajectory choices that lead to 4- or 5-hour eclipses. There are also many choices that result in maximum eclipses as low as 2 or 3 hours, well within the mission constraint. 

\section{Fully Integrated Mission Trajectory}
The aforementioned analyses were necessary to gain insight into the orbit regime (i.e., to narrow the space of desirable orbits to something manageable) and to verify the robustness of the P/2-HEO orbit. With this knowledge in hand, STK's Astrogator module was used to design 27 candidate mission trajectories (one for each day of the lunar cycle), using in particular the guidance of Fig.~\ref{fig:PLEPMisalignment} to provide the Astrogator targeter with orbits that should yield desirable behavior throughout the mission life. The STK-derived trajectories are fully patched from the launch through a long-duration (4-year) simulation of the mission orbit. They include all the necessary (targeted) maneuvers to maintain phasing-orbit perigee above 600 km, to raise phasing-orbit apogee, to target the lunar flyby, and to inject into the mission orbit at PLEP. Stationkeeping $\Delta V$ is not required for the P/2 HEO orbit.

Figures~\ref{fig:MinPerigeeSTK} and~\ref{fig:MinIncSTK} show the results of these fully-patched missions compared to the TRACE simulations, whose initial conditions were calculated with the lower fidelity analytical methods. The differences are expected, but it is clear that the analytical guidance was vital to choosing the correct orbital parameters that satisfy the mission constraints.

The total mission $\Delta V$ for the 27 candidate trajectories as a function of lunar flyby day of the month appears in Figure~\ref{fig:STKdv}. Also included is the $\Delta V$ for a hypothetical large observatory mission. Assuming a wet mass of ~7000 kg, an Atlas V has the capability to inject the spacecraft into an initial phasing orbit with apogee of 100,000 km. The additional $\Delta V$ required to complete the apogee raise maneuvers to 400,000 km (as compared to the TESS initial apogee of 250,000 km) is 208 m/s. The $\Delta V$ also includes the phasing-orbit apogee-raise maneuvers and the PAM burn to inject into the mission orbit. Each trajectory budgets 8 m/s for maintaining phasing-orbit perigee, 28 m/s for launch dispersions, and 25 m/s for trajectory correction maneuvers, even if they were not modeled in the simulation. As long as the flyby epoch is the same for the TESS DRM and the large spacecraft mission, the $C_{\textrm{\tiny phas}}$ at the time of flyby is about the same, and the PAM $\Delta V$ should be roughly equal, so the whole plot simply shifts up by 208 m/s.

\begin{figure}[htb]
  \begin{minipage}[t]{0.483\linewidth}
    \centering
    \includegraphics[width=2.5in]{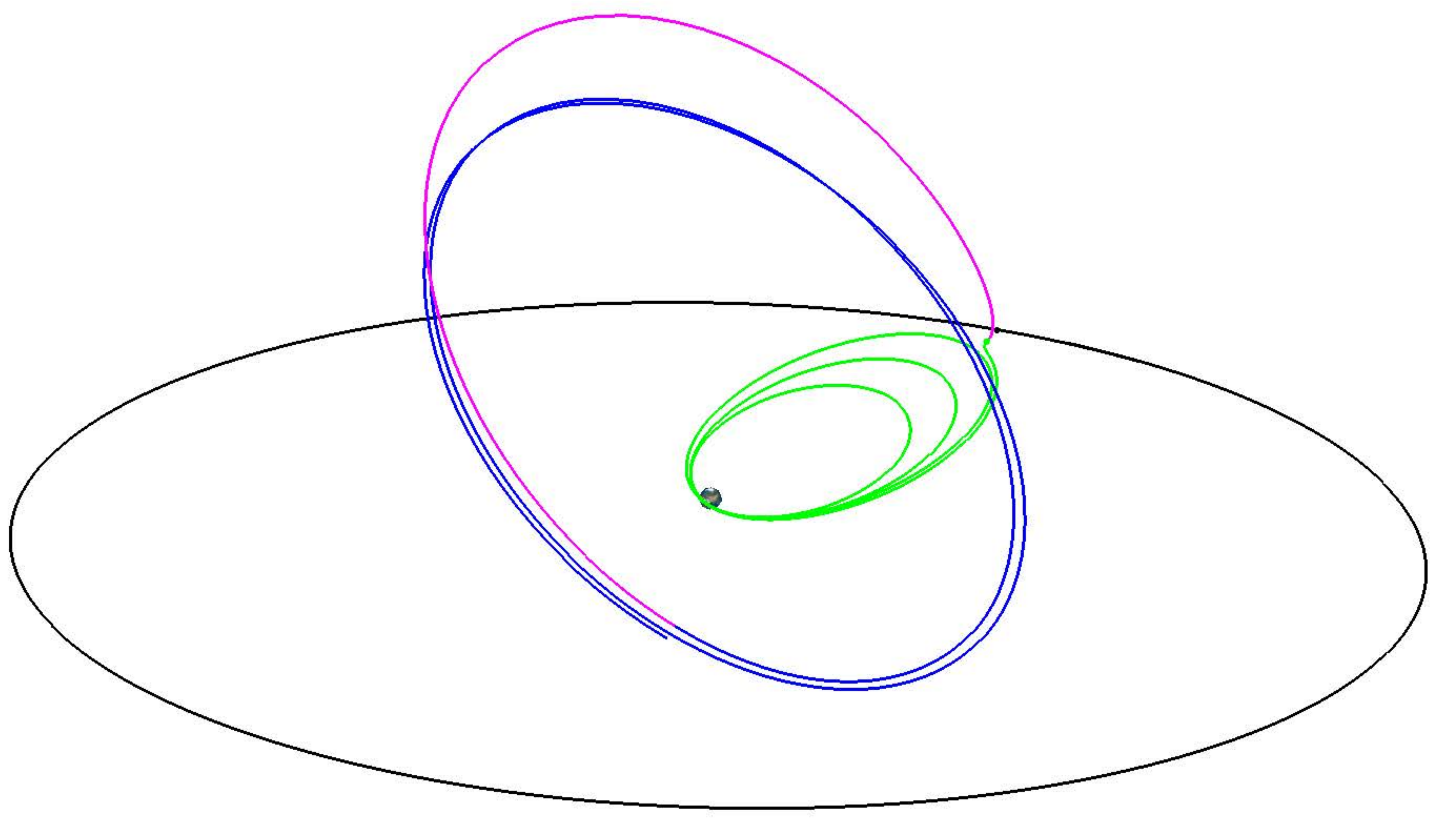}
    \caption{The TESS DRM showing the phasing orbits (green), the transfer orbit (purple), and the final mission orbit (blue). The Moon's orbit is black.}
    \label{fig:DRM}
  \end{minipage}%
  \hspace{0.5cm}
  \begin{minipage}[t]{0.483\linewidth}
    \centering
    \includegraphics[width=2.5in]{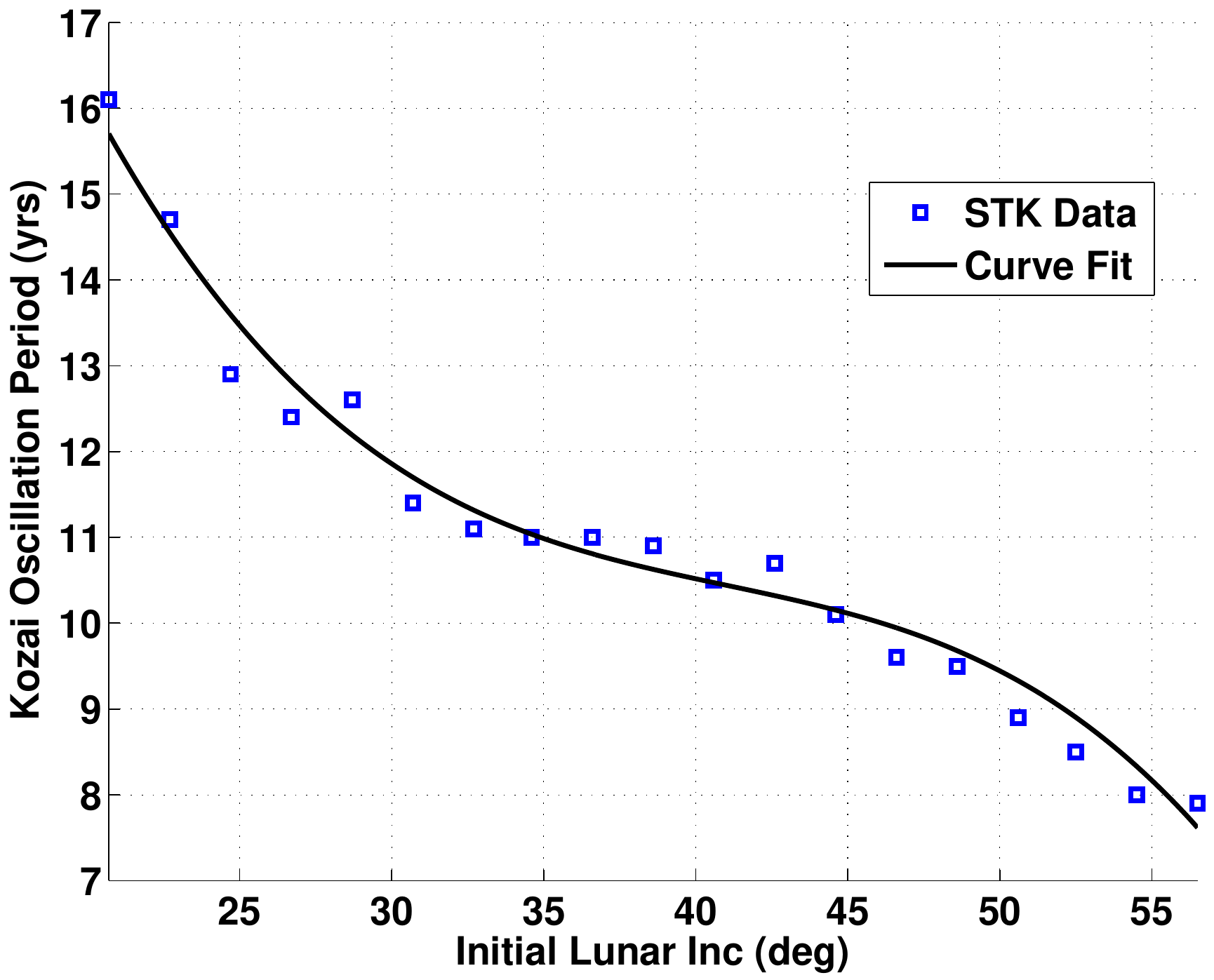}
    \caption{Period of Kozai oscillation versus initial lunar inclination.}
    \label{fig:TESSkozaioscil}
  \end{minipage}
\end{figure}

The DRM selected for TESS corresponds to the lunar flyby on day 9 of the June 2017 lunar cycle, which is the minimum $\Delta V$ trajectory from the set of candidates in Fig.~\ref{fig:STKdv}. The DRM satisfies the minimum-perigee constraint (cf.~Fig.~\ref{fig:MinPerigeeSTK}) and the eclipse constraint. An STK illustration of the mission trajectory up to the final mission orbit appears in Figure~\ref{fig:DRM}.  The figures show (in chronological order) the phasing orbits (green), the transfer orbit (purple), and the final mission orbit (blue). The Moon's orbit is black.

Some key parameters of the 4-year mission-orbit simulation results appear in Table~\ref{tab:TESSDRMparams}.  The lunar flyby date of 6 July 2017 is the epoch that drives the mission design, as the launch date is a function of the lunar flyby approach strategy (i.e., duration of phasing orbits). For TESS, the 3.5 phasing orbits preceding the DRM lunar flyby define a launch date of 6 June 2017. The DRM maintains the perigee variance within the constraint bounds, and has a maximum eclipse of 4.4 hours in duration, safely within the 6-hour limit. The hypothetical 7,000 kg spacecraft launched into an initial 100,000 km apogee phasing orbit on an Atlas V previously discussed produces similar results. Although the total $\Delta V$ requirement is quite different, as long as the flyby epoch is the same, the only difference between the two missions is the size and duration of the phasing orbits, which means the time from launch to lunar flyby is different. The difference in lunisolar perturbations causes $C_{\textrm{\tiny phas}}$ to vary somewhat, but because most of the time is spent in the final few 400,000 km apogee phasing orbits in both missions, the effect is very small. The post-flyby orbits for both missions are nearly identical.
The $\Delta V$ budget in Table~\ref{tab:DRMdv} includes the same margin for unforeseen maneuvers included in Fig.~\ref{fig:STKdv}. For the TESS DRM simulation (and the hypothetical ~7000 kg spacecraft launched on an Atlas V for the same flyby epoch), no dispersions or TCMs were modeled, and no phasing-orbit perigee raise maneuvers were required. Many of the candidate missions in the June 2017 lunar cycle do require additional $\Delta V$ for such maneuvers, and the budget in Table 2 is sufficient for those cases. 

\begin{table}
\centering
\makebox[0pt][c]{\parbox{1.2\textwidth}{%
    \begin{minipage}[b]{0.5\hsize}\centering
        \caption{Key Parameters of the TESS DRM}
\begin{tabular}{lc}
\hline
\hline
Parameter & STK Value \\
\hline
Launch Epoch & 6 June 2017 14:16:25.6 \\
Lunar Flyby Perilune Epoch & 6 July 2017 09:48:57.8 \\
Lunar Flyby Altitude, km & 7,925 \\
Ascending/Descending @ Flyby & Ascending \\
Minimum Perigee, $R_E$ & 8.4 \\
Maximum Perigee, $R_E$ & 21.2 \\
Min Ecliptic Inclination, deg & 20.9 \\
Maximum Eclipse, hr & 4.4 \\
\hline
\hline
\end{tabular}
        \label{tab:TESSDRMparams}
    \end{minipage}
    \hfill
    \begin{minipage}[b]{0.5\hsize}\centering
\caption{$\Delta V$ budget for the TESS DRM and a similar large spacecraft mission.}
\begin{tabular}{p{0.9in}p{0.6in}p{0.5in}p{0.5in}}
\hline
\hline
Event & TESS Allocated, m/s & TESS STK, m/s & Atlas V, ~7000 kg, m/s \\
\hline
LV Dispersions & 28 & N/A & N/A \\
Phasing-Orbit Perigee Raises & 8 & 0 & 0 \\ 
Phasing-Orbit Apogee Raises & 54 & 54 & 262 \\
TCMs & 25 & N/A & N/A \\
PAM & 95 & 71 & 71 \\
Stationkeeping & 0 & 0 & 0 \\
\textbf{Total} & 210 & 125 & 333 \\
\hline
\hline
\end{tabular}    
        \label{tab:DRMdv}
    \end{minipage}%
}}
\end{table}

\begin{figure}[t]
\hspace*{\fill}
\subfigure[Perigee versus lunar inclination ($K = 0.65$).]{\includegraphics[width=2.3in]{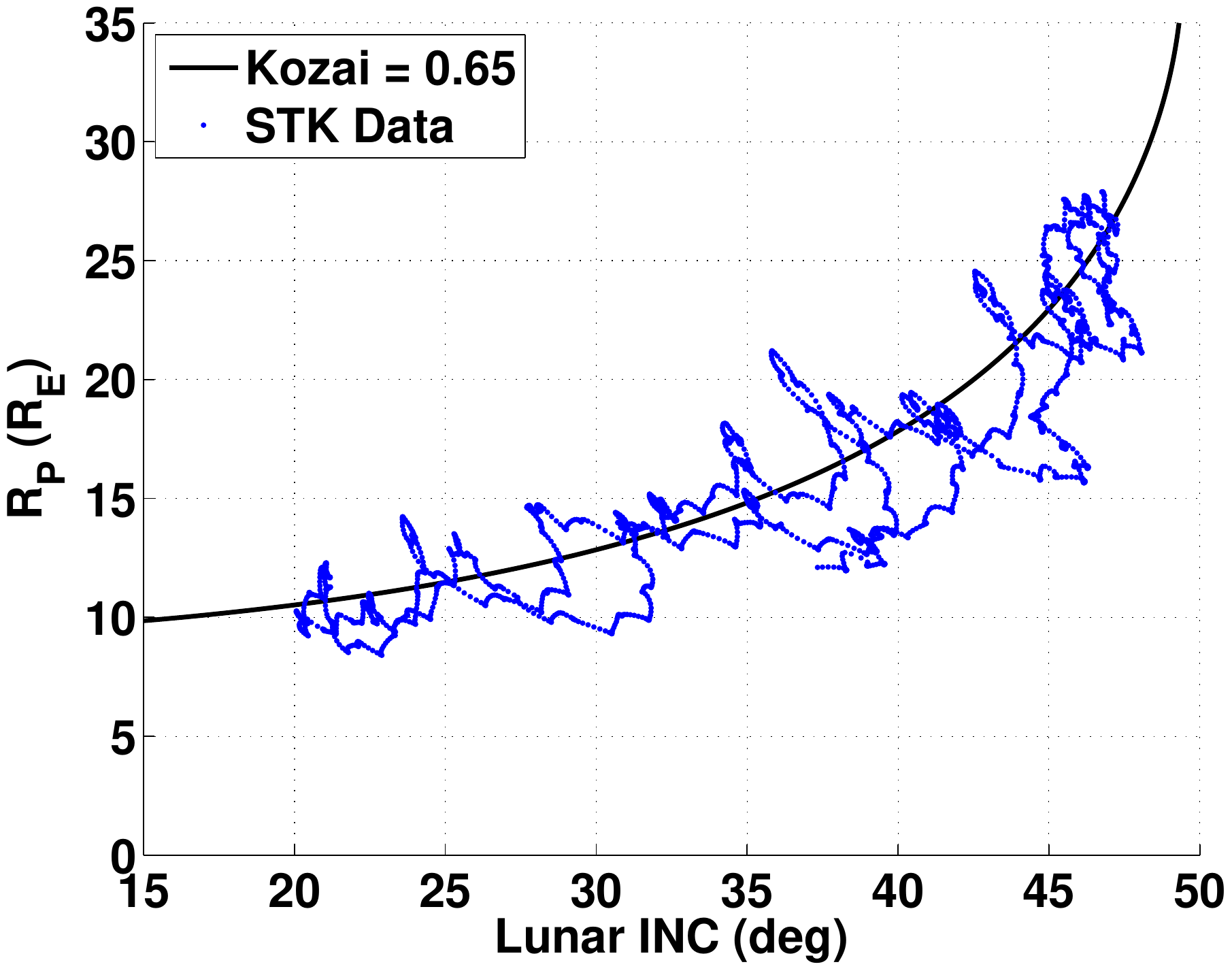}\label{fig:TESSrpvsinc}}
\hfill
\subfigure[Lunar inclination versus AOP.]{\includegraphics[width=2.3in]{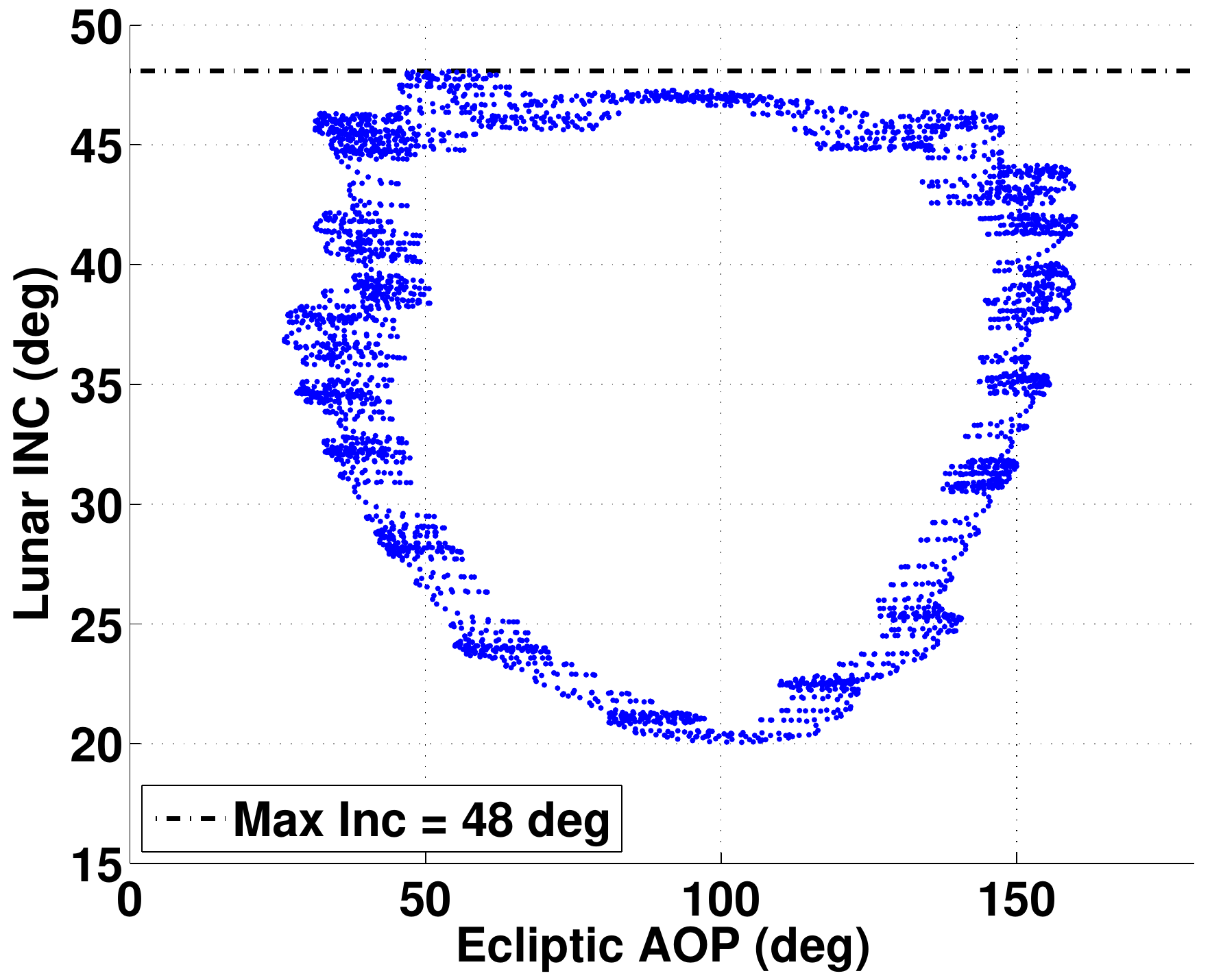}\label{fig:TESSincvsaop}}
\hspace*{\fill}
\caption{The behavior of the TESS DRM's perigee, inclination, and argument of perigee as a function of time.}
\label{fig:TESSkozaicomp}
\end{figure}

Figure~\ref{fig:TESSkozaioscil} illustrates the relationship of the P/2-HEO Kozai cycle period to initial lunar inclination. Detailed numerical analysis indicates that the DRM (initial inclination of 38 degrees) has a Kozai period of ~11 years. The Kozai cycle, shown previously in Fig.~\ref{fig:TESSorbels} for the TESS DRM, varies from 8 to 16 years depending on initial inclination. Close examination of Fig.~\ref{fig:TESSorbels} indicates that the long term variations are related to Kozai and the shorter term variations (on the order of 6 months) are due to solar perturbations. Figure~\ref{fig:TESSrpvsinc} demonstrates the relationship of DRM mission perigee and lunar inclination for one Kozai cycle as defined by Eq.~\ref{eq:Kozai}, where $K = 0.65$. Figure~\ref{fig:TESSincvsaop} illustrates the maximum lunar inclination for one Kozai cycle.

The TESS DRM is a point-design P/2-HEO trajectory that satisfies all the mission constraints.  Many other such point designs certainly exist, both during this and other lunar cycles.  The value of the analyses performed here is that the sensitivity of the perigee variance and eclipse duration to changes in the lunar flyby date and initial conditions is now understood, and a well-informed, formulaic approach is available to reproduce a similar trajectory subject to different operational constraints.

\section{Conclusions}
Although the trade space for the P/2-HEO is large in general, it has been shown that the application of typical mission-related constraints---including maximum eclipse durations, minimum and maximum perigees, and target values of the Kozai constant to ensure desired orbit variability---reduces the space to only a handful of variables. Much of this space can be explored analytically, greatly reducing computational overhead and providing unique insight into the P/2-HEO that would be unavailable if the problem were approached from a strictly numerical perspective. The analytical guidance for finding desirable orbital elements of the P/2-HEO was verified by high-fidelity numerical integrations and subsequently used to seed high-fidelity software that produced point designs and, ultimately, the TESS DRM along with daily trajectories for 27 days in the June 2017 lunar cycle. Although TESS was the motivating example for this paper, the P/2-HEO is a competitive option for a mission of any size. This unique orbit provides the stable thermal and attitude-control environment and the unobstructed views of the sky essential for high-quality scientific measurements, while being reachable for considerably less $\Delta V$ than traditional alternatives.

\section{Acknowledgments}
The authors would like to thank the following individuals for their contributions to the detailed analysis of the TESS P/2-HEO presented in this paper. Jose Guzman and Robert Lockwood of the Orbital Sciences Corporation first suggested the P/2-HEO as a strong candidate for TESS, and Robert Lockwood participated in and reviewed all of the analysis, as well as offered his experience from IBEX. Chad Mendelsohn of the Goddard Space Flight Center performed independent verification of the detailed mission orbits analysis. Finally, Dave Bearden, Matt Hart, Debra Emmons, and Warren Goda of The Aerospace Corporation obtained the research funds that enabled this work.

\bibliographystyle{AAS_publication}   
\bibliography{references}   

\end{document}